\documentclass[10pt,reprint]{revtex4-2}
\usepackage[english]{babel}
\usepackage{amsmath,amssymb,amsthm}
\usepackage{braket}
\usepackage{enumerate}
\usepackage{wrapfig}
\usepackage{subcaption}
\usepackage{xcolor,color,tikz}
\usetikzlibrary{positioning,shapes.geometric,decorations.markings,decorations.pathmorphing,arrows,knots,hobby,math}
	\tikzset{
		state/.style={circle,draw,minimum size=6ex},
		arrow/.style={-latex, shorten >=1ex, shorten <=1ex}}
\usepackage[]{todonotes} 
\usepackage{hyperref}

\definecolor{light-gray}{gray}{0.85}

\theoremstyle{plain}

\theoremstyle{definition}
\newtheorem{definition}{Definition}[section]

\DeclareMathOperator{\tr}{tr}

\DeclareMathOperator{\real}{Re}
\DeclareMathOperator{\imag}{Im}
\DeclareMathOperator{\diff}{diff}

\DeclareMathOperator{\arcsinh}{arcsinh}

\DeclareMathOperator{\PSL}{PSL}
\DeclareMathOperator{\SL}{SL}
\DeclareMathOperator{\SO}{SO}
\DeclareMathOperator{\oo}{O}

\definecolor{color1}{RGB}{0,80,155}
\definecolor{color2}{RGB}{125,125,125}
\definecolor{color3}{RGB}{0,117,223}

\date{\today}

\begin{document}

\title{Holographic networks for $(1+1)$-dimensional de Sitter spacetime}

\author{Laura Niermann}
\author{Tobias J.\ Osborne}
\affiliation{Institut f\"ur Theoretische Physik, Leibniz Universit\"at Hannover, Appelstr. 2, 30167 Hannover, Germany}

\begin{abstract} 
Holographic tensor networks associated to tilings of $(1+1)$-dimensional de Sitter spacetime are introduced. Basic features of these networks are discussed, compared, and contrasted with conjectured properties of quantum gravity in de Sitter spacetime. Notably, we highlight a correspondence between the quantum information capacity of the network and the cosmological constant.
\end{abstract}

\maketitle

\section{Introduction }
The duality between quantum gravity theories in anti-de Sitter spacetime (AdS) in the semiclassical limit and strongly interacting  conformal field theories (CFT) -- the AdS/CFT correspondence -- has provided many deep insights into the structure of quantum gravity. Since the original incarnation \cite{maldacena_large-n_1999,maldacena_large_1998} of this  duality there has been a Cambrian explosion of studies exploring holographic correspondences in a variety of settings. These lines of enquiry have ramified throughout high energy physics and quantum many body theory, branching out from the initial works of Gubser, Klebanov, and Polyakov \cite{gubser_gauge_1998}, and Witten \cite{witten_anti_1998}, and providing an intricate yet coherent understanding of quantum gravity in AdS.

Observational evidence \cite{schmidt_high-z_1998}, however, strongly favours the hypothesis that we live in a universe which is asymptotically de Sitter (dS). It is therefore of critical importance to transfer our holographic knowledge and expertise concerning quantum gravities formulated in AdS to the dS setting. This process has been ongoing but progress has been slow in comparison to the AdS case. Key works on quantum gravity in dS include early papers of Banks \cite{banks_cosmological_2001}, Bousso \cite{bousso_covariant_1999,bousso_holography_1999,bousso_positive_2000}, Witten \cite{witten_quantum_2001}, and Balasubramanian \emph{et al} \cite{balasubramanian_deconstructing_2001}. The application of holographic ideas was then initiated in the crucial paper of Strominger \cite{strominger_ds/cft_2001}. Since then there have been a steady stream (see, e.g., \cite{balasubramanian_mass_2002, anninos_higher_2016} and references therein) of works investigating this duality. Progress has been impeded by a multitude of obstructions, not least of which is that the observables of dS, living at timelike infinity, are radically different to those of AdS which live at spatial infinity.

While a complete quantum gravity theory in dS in $(3+1)$ dimensions is still largely out of reach, there has been considerable recent progress in understanding dS in low dimensions. In particular for the case of Jackiw-Teitelboim theory \cite{jackiw1985lower,jackiw1992gauge,teitelboim1983gravitation} in $(1+1)$ dimensions there has been recent progress in understanding the structure of the Hilbert space and unitarity of the theory \cite{maldacenaTwoDimensionalNearly2021,Cotler2020_LowdimdS,cotlerEmergentUnitaritySitter2019}. Although there are no bulk gravitons for such theories they do share the property that all the observables live at the temporal boundaries. 

From the perspective of the present work, however, the most remarkable feature of quantum gravity in dS is the proposed \emph{finite-dimensionality} of its kinematical Hilbert space. This surprising and counterintuitive conclusion was argued persuasively by Bousso \cite{bousso_covariant_1999,bousso_holography_1999,bousso_positive_2000} (building on earlier work of Banks \cite{banks_cosmological_2001}) and points to a deep connection -- the \emph{$\Lambda - N$ correspondence} -- between the cosmological constant and the dimensionality of Hilbert space. If true, this correspondence suggests that tools arising in quantum information theory, particularly in the theory of quantum entanglement, may be useful in exploring quantum gravity in dS. 

There has already been very active interest in exploring connections between quantum gravity and quantum information theory. This direction originates, in part, from works of {Ryu} and Takayanagi \cite{ryu_holographic_2006,rangamani_holographic_2017}, who expressed the quantum entanglement entropy of a region in the boundary CFT in terms of the area of a specific bulk minimal surface. These ideas have been considerably expanded and developed since then commencing with the proposals of Van Raamsdonk \cite{van_raamsdonk_building_2010} and Swingle \cite{swingle_entanglement_2012} and culminating in the ER=EPR proposal of Susskind and Maldacena \cite{maldacena_cool_2013}.  The tools of quantum entanglement and Hamiltonian complexity theory are now also being used to profitably explore the quantum dynamics of black holes \cite{harlow_jerusalem_2016,brown_complexity_2016,hayden_BlackHoles_2007}.

Another central tool in quantum information theory, quantum error correction (QEC), has also played an important role in understanding bulk/boundary correspondences. This was initiated by Almheiri, Harlow, and Dong \cite{almheiri_bulk_2015} who argued that bulk local operators should manifest themselves as logical operators on subspaces of the boundary CFT of $\text{AdS}_3$ spacetime. This paper was an important inspiration for the construction of a remarkable toy model of holographic duality known as the \emph{holographic code} \cite{pastawski_holographic_2015}, a discrete model of the kinematical content of the AdS/CFT correspondence built from special tensors coming from QEC known as \emph{perfect tensors}.  Holographic codes have proved to be a very helpful rosetta stone for quantum information theorists and high energy physicists and there has been considerable progress investigating and generalising  holographic codes \cite{pastawski_holographic_2015,hayden_holographic_2016,bao_consistency_2015,yang_bidirectional_2016,bhattacharyya_exploring_2016,may_tensor_2017}
and perfect tensors \cite{goyeneche_absolutely_2015,enriquez_maximally_2016,raissi_optimal_2018,li_invariant_2017,peach_tensor_2017,donnelly_living_2017}. Exploiting powerful results \cite{jones_unitary_2014,jones_no-go_2018} of Jones on unitary representations of a discrete analogue of the conformal group known as Thompson's group $T$ \cite{cannon_introductory_1996,belk_thompsons_2007}, the kinematical holographic state has been upgraded to yield a full dynamical toy model of the AdS/CFT correspondence.

Our paper represents an initial exploratory step towards transferring recent techniques in the study of holographic codes to the de Sitter setting. The primary motivation for the current investigation lies in the observation that tensor networks such as tree-tensor networks and MERA have a causal structure which naturally matches that of spacetimes with $\Lambda > 0$ \cite{swingle_entanglement_2012,czech_integral_2015,beny2013causal,czech_tensor_2016,qi_exact_2013,bao_sitter_2017,milstedGeometricInterpretationMultiscale2018a}. A second motivation is the argued finite-dimensionality of the de Sitter gravitational Hilbert space. To concretise these ideas we study the simplest possible case of  de Sitter spacetime in $(1+1)$-dimensions by proposing certain holographic tensor networks from discretisations of de Sitter spacetime as a microscopic kinematical model. In contrast to the AdS case it turns out that these networks are better understood as defining a propagator or, more precisely, a \emph{partial isometry}, from the past boundary to the future boundary on a finite-dimensional \emph{physical} subspace of an infinite-dimensional boundary Hilbert space. We then identify a fundamental correspondence between the quantum information capacity of the network and the cosmological constant. Finally, discrete analogues of diffeomorphisms are introduced via Thompson's group $T$, and their action on the physical subspace  characterised. The tensor networks we propose here have a treelike structure reminiscent of the proposals involving $p$-adic numbers \cite{gubserPAdicAdSCFT2017b,gubserAdicVersionAdS2017} and resemble the eternal symmetree proposal \cite{harlowTreelikeStructureEternal2012a}. Our proposal differs from constructions involving $p$-adic numbers, however, because the action of relevant symmetry groups (i.e., in our case Thompson's group $T$) is not equivalent. 

\section{De Sitter spacetime}
In this section the basic properties of de Sitter spacetime are introduced and reviewed. Readers familiar with de Sitter spacetime solutions may safely skip this section.

In 1917 de Sitter reported two solutions for Einstein's field equations \cite{de_sitter_relativity_1917,de_sitter_curvature_1917}, namely maximally symmetric spacetimes with constant positive (respectively, negative) curvature. In the case of constant curvature the Riemann tensor $R^\rho_{\sigma\mu\nu}$, is already determined by the Ricci scalar $R$. Since \cite{hawking_large_1973} the Ricci scalar $R$ needs to be constant for a spacetime with constant curvature, de Sitter and anti-de Sitter spacetimes are solutions of the field equations for empty spacetime with cosmological constant $\Lambda = \frac14 R$.
Spacetimes having {constant} positive curvature $R>0$ are called \emph{de Sitter} (respectively, for negative curvature $R<0$, \emph{anti-de Sitter}). 

Based on observations in 1998 concerning type Ia supernova \cite{schmidt_high-z_1998} it has been hypothesised that the cosmological constant $\Lambda$ of our universe is small and positive. Thus, de Sitter spacetime could be a natural limiting spacetime for our universe (see, e.g., \cite{suneeta_notes_2002}). These observations considerably motivate us to adapt or transfer the understanding of quantum gravity for anti-de Sitter spacetime to the de Sitter case most relevant for our universe.

The most straightforward way to define $d$-dimensional de Sitter spacetime dS$_d$ with one temporal and $d-1$ spatial dimensions is by describing it as a hypersurface embedded in $(d+1)$-dimensional Minkowski spacetime: it is then a single-sheeted hyperboloid in Minkowski spacetime. The points of $d$-dimensional de Sitter spacetime fulfil the hyperboloid condition:
\begin{equation}
	-x_0^2 + x_1 ^2 + \cdots + x_d^2 = \ell^2. \label{eq:hyperboloid}
\end{equation}
The parameter $\ell$ on the RHS is the \emph{de Sitter radius}, which describes the size of space at the time where its size is minimal. We parametrise the temporal coordinate $x_0$ so that the radius is smallest at $x_0=0$. If not otherwise specified, we assume the de Sitter radius is given by $\ell=1$.
The metric of $d$-dimensional de Sitter spacetime is given by 
\begin{equation}
	\mathrm{d}s^2 = -\mathrm{d}x_0^2 + \mathrm{d}x_1 ^2 + \cdots + \mathrm{d}x_d^2 \label{eq:metric}.
\end{equation}

\begin{figure}
	\begin{center}
		\includegraphics[page=1]{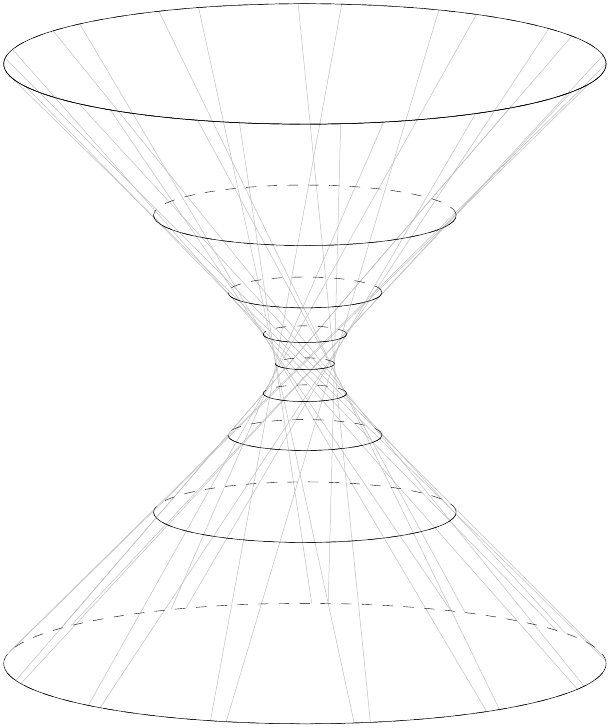}
	\end{center}
	\caption{dS$_2$ embedded in Minkowski spacetime (grey lines are null geodesics, black circles are constant time slices) }
	\label{fig:dS_2}
\end{figure}
For the rest of this paper we focus on the $(1+1)$-dimensional case, depicted in Fig.~\ref{fig:dS_2} as a single-sheeted hyperboloid  in Minkowski spacetime. The geodesics used in the figure are reviewed in Appendix \ref{app:geodGlobal}. It is worth noting that $d$-dimensional de Sitter spacetime has \emph{two} temporal boundaries and 
\emph{no} spatial boundary. (AdS has, by contrast, a single spatial boundary.) The temporal boundaries lie in the temporal future (respectively, past) infinity. This structure has far-reaching consequences for the formulation of quantum gravity. 

\subsection{Null geodesics in two-dimensional de Sitter spacetime}\label{sec:nullgeodesics}
Geodesics are an important tool to describe and analyse the structure of spacetime. The geodesics we are most interested in here are null geodesics, which describe the propagation of light rays. Two-dimensional de Sitter hypersurface is ruled by null geodesics (there are two distinct straight lines running through every point in the surface). 

The null geodesics in de Sitter spacetime may be constructed from null geodesics in $(d+1)$-dimensional Minkowski spacetime by imposing the hyperboloid constraint Eq.~(\ref{eq:hyperboloid}):
\begin{equation}
	x(s) 
	= \begin{pmatrix} x_0 \\ x_1 \\ x_2 \end{pmatrix} 
	= \begin{pmatrix}s \\ u \pm vs \\ v \mp us\end{pmatrix}
	, \quad s\in\mathbb{R}, \, u^2+v^2 = 1. \label{eq:geodesic}
\end{equation}
The condition $u^2+v^2=1$  can be observed by writing the hyperboloid condition in matrix form as 
\begin{equation}
	x^T\eta x = 1,\quad
	\text{with }\,
	\eta = \text{diag}(-1,1,\cdots,1).
	\label{eq:hyperboloidMatr}
\end{equation}
We now see that the constraint  $u^2+v^2=1$ on the parameters of the null geodesic $x(s)$ arises as follows:
\begin{align*}
	&x^T(s) \eta x(s) = -s^2 + (u\pm vs)^2 + (v\mp us)^2\\
	&= -s^2 + (u^2+v^2) + s^2 (u^2+v^2)
	= 1.
\end{align*}

Each null geodesic is specified by its intersection point with the time slice $x_0=0$ and a sign.  The sign distinguishes between the two different classes of null geodesics, which we call \emph{anticlockwise}- and \emph{clockwise}-pointing null geodesics, respectively. This is depicted in Fig.~\ref{fig:dS_2}, where one observes the direction of propagation around the spatial coordinate of the null geodesic on the hyperboloid is either clockwise or anticlockwise.

The parameters for the clockwise and anticlockwise pointing null geodesics only differ by a sign. This is why in the sequel we usually only consider the anticlockwise case as a representative of null geodesics in general. The calculations are very similar for clockwise-pointing null geodesics. 

\subsection{Global and conformal coordinates} 
It is convenient to introduce a coordinatisation to easily work with de Sitter spacetime. The most straightforward coordinate system for $d$-dimensional de Sitter spacetime is furnished by \textit{global coordinates}, so named because they describe the entire single-sheeted hyperboloid embedded in Minkowski spacetime (for further details on coordinate systems for de Sitter spacetime see, e.g., \cite{spradlin_sitter_2002} for further details). Global coordinates are defined via
\begin{align}\begin{split}
		x_0 &= \sinh(\tau) \\
		x_j &= \omega_j \cosh(\tau), \quad j = 1,2, \ldots, d,
\end{split}\end{align}
where $\tau \in \mathbb{R}$ is the temporal variable and the angular variables $\omega_i$ are defined according to
\begin{align*}
	\omega_1 &= \cos(\theta_1), \\ 
	\omega_2 &= \sin(\theta_1)\cos(\theta_2), \\
	&\vdots \nonumber\\
	\omega_{d-1} &= \sin(\theta_1)\cdots \sin(\theta_{d-2})\cos(\theta_{d-1}),\\
	\omega_d &= \sin(\theta_1)\cdots \sin(\theta_{d-2})\sin(\theta_{d-1}), 
\end{align*}
where $0\le \theta_j < \pi$ for $1\le j < d-1$ and $0\le \theta_{d-1} < 2\pi$.\\

For the two-dimensional case, this simplifies somewhat and only one angle variable $\theta$ with $0\le \theta < 2\pi$ is necessary to define global coordinates (here $\theta=0$ and $\theta=2\pi$ are identified):
\begin{align}\label{eq:globalcoord}\begin{split}
		x_0 &= \sinh(\tau), \\
		x_1 &= \cos(\theta)\cosh(\tau), \\
		x_2 &= \sin(\theta)\cosh(\tau).
\end{split}\end{align}
The de Sitter metric Eq.~(\ref{eq:metric}), parametrised in global coordinates, takes the form
\begin{align}
	\mathrm{d}s^2 = -\mathrm{d}\tau^2 + \cosh^2(\tau)\, \mathrm{d}\theta^2.
\end{align} 
(This is reviewed in Appendix~\ref{app:metricGlobal}.)
Null geodesics may be parametrised according to
\begin{align}
	\begin{pmatrix}
		\theta(t) \\ \tau(t) 
	\end{pmatrix} = \begin{pmatrix}
		\pm \arctan t + \theta_0 \\ \arcsinh t 
	\end{pmatrix},\qquad t\in \mathbb{R}.
\end{align}

It is useful to introduce a new temporal variable to describe the causal structure of de Sitter spacetime. This also makes the temporal boundaries of de Sitter spacetime somewhat more amenable to direct study. The corresponding new coordinates are called \textit{conformal coordinates}. This coordinate system exploits the same spatial variables as global coordinates, however, a new temporal variable $T$ is introduced via 
\begin{equation}
	\cos T= \frac{1}{\cosh \tau },\quad \text{where } -\frac{\pi}{2} < T < \frac{\pi}{2}. \label{eq:ConformalCoordinates}
\end{equation}
This coordinate transformation maps the temporal infinities in global coordinates ($\tau=\pm\infty$) to $T=\pm \frac{\pi}{2}$, respectively. Accordingly, the temporal variable in conformal coordinates is always finite, even at temporal infinity.
\begin{figure}
	\begin{flushright}
		\includegraphics[page=2]{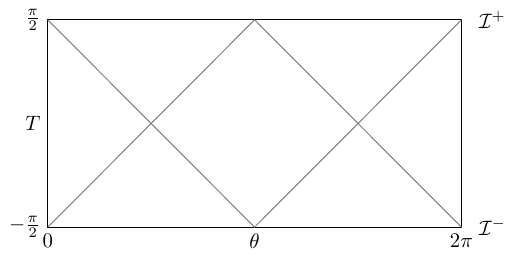}
		\caption{dS$_2$ in conformal coordinates where $\theta=0$ and $\theta=2\pi$ are identified (grey lines are null geodesics).}
		\label{fig:dS_conformal}
	\end{flushright}
\end{figure}  
In conformal coordinates a point in two-dimensional de Sitter spacetime may be parametrised with the spatial coordinate $\theta$ and the temporal coordinate $T$: \begin{equation}
	p=(\theta,T).
\end{equation}
In this way $(1+1)$-dimensional de Sitter spacetime may be represented by a rectangle (see Fig.~\ref{fig:dS_conformal}). Geodesics in conformal coordinates are straight lines tilted at $45^\circ$, just as they are in Minkowski spacetime (this is reviewed in Appendix \ref{app:geodesicProof}). A null geodesic in conformal coordinates may be parametrised via:
\begin{align}
	\begin{pmatrix}
		\theta(s) \\ T(s)
	\end{pmatrix} = \begin{pmatrix}
		\theta_0 \pm s \\ {s}
	\end{pmatrix},  \qquad -\frac{\pi}{2}<s<\frac{\pi}{2}.\label{eq:geodConfCoord}
\end{align} 
A key advantage of the conformal coordinate system is that it allows us to transparantly understand the causal structure of de Sitter spacetime. Further, many calculations are easier to perform on a rectangle, as opposed to a hyperbolic surface. 

\subsection{Causal structure and causal diamonds} 
\label{sec:CausalDiamonds}

In this section causality relations and the accessibility of information in de Sitter spacetime are considered. 
To this end we introduce \textit{causal diamonds}, which are special subsets of de Sitter spacetime. 

In order to explain the physical relevance of causal diamonds consider an experiment moving along a worldline commencing at a spacetime location $p$ and concluding at location $q$. A causal diamond is then that subset of de Sitter spacetime where the information influencing --- and influenced by --- the experiment between point $p$ and point $q$ is accessible to an external observer. For a review see \cite{hawking_large_1973}.

\begin{figure}\begin{center}
		\includegraphics[page=3]{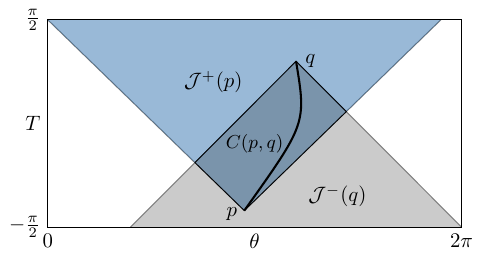}
		\caption{A causal diamond is the intersection of the causal future of point $p$ and the causal past of point $q$.}
		\label{fig:causalDiamond}
\end{center}\end{figure}

There are two restrictions that have to be fulfilled for a point to lie within the causal diamond defined by the points $p$ and $q$, namely, that the information needs to lie within the causal future $\mathcal{J}^+(p)$ of point $p$ and within the causal past $\mathcal{J}^-(q)$ of point $q$:
\begin{equation}
	C(p,q)=\mathcal{J}^+(P) \cap \mathcal{J}^-(q).
\end{equation}
In this way a causal diamond $C(p,q)$ is defined only by the points $p$ and $q$, and not by the details of the experiment's worldline. See Fig.~\ref{fig:causalDiamond} for an illustration. 
Note that any continuous non-spacelike worldline from point $p$ to point $q$ lies within the causal diamond.

The largest influence a local experiment can have on observers within de Sitter spacetime is then found by sending $p$ to the {temporal} past infinity and $q$ to the {temporal} future infinity:
\begin{equation*}
	p \rightarrow x_-\in \mathcal{I}^- \qquad q \rightarrow x_+\in \mathcal{I}^+.
\end{equation*}
Everything an observer can do and observe is thus determined by the causal diamond $C(x_-,x_+)$ defined by just these two points $x_\pm$ lying within the temporal boundaries. Such a causal diamond is of maximal size if the defining points have the same $\theta$ coordinate. It is impossible to cover the entirety of de Sitter spacetime using just one causal diamond; no single observer can acquire information about an arbitrary point in de Sitter spacetime, even in principle!  Contrast this with Minkowski spacetime: here any local experiment may, in principle, influence, and be influenced by, any point within the spacetime. Causal diamonds are a central building block in section \ref{sec:tessellation}, where a tessellation for a two-dimensional de Sitter spacetime is constructed.

A striking consequence of the causal structure of de Sitter spacetime, as emphasised by Witten \cite{witten_quantum_2001}, is that there is no global positive conserved energy quantity. This is a consequence of the absence of a global timelike Killing vector field (although it is possible to define local timelike Killing vector fields), so there is no global generator of time-translation symmetry. Therefore it is at best questionable -- and likely impossible -- whether one can define a \emph{unitary} time-translation operator with a corresponding Hamiltonian generator on the quantum gravity Hilbert space. This does however leave open the possibility that time translation might be implemented quantum mechanically via a dissipative process described via a completely positive map.

If we take all of these observations at face value we are forced to accept that the observables of quantum gravity in dS live on $\mathcal{I}^{\pm}$. This is in stark contrast to the AdS case where we have a wealth of observables on the spatial boundary which we may identify with observables of a conformal field theory on the holographic boundary. These features (or bugs, depending on your point of view) make reasoning about quantum gravity in $\text{dS}$ rather different to its AdS counterpart.

The -- here assumed -- impossibility of defining a Hamiltonian for de Sitter spacetime is one core reason why in this paper we consider the quantum mechanics of de Sitter spacetime only at its boundaries. The model we construct is defined on a Hilbert space attached to temporal infinity. As a result, our model gives us only indirect insight, which must be holographically reconstructred, on events within de Sitter spacetime at finite times.

\section{A holographic network for de Sitter spacetime}
There are profound reasons to believe that a quantum theory of gravity can be formulated holographically \cite{van2017lectures}, a programme which has been pursued most profitably in the AdS case. Our goal in this section, motivated by the proposal of \cite{pastawski_holographic_2015} for AdS, is to commence the exploration of toy holographic formulations of de Sitter spacetime using tensor networks associated to tessellations in as direct a way as possible. In this way we construct a holographic toy model for two dimensional de Sitter spacetime. It should be noted, however, that because de Sitter spacetime has two temporal boundaries the model we construct differs drastically from the one presented in \cite{pastawski_holographic_2015}, and it does not directly fit the definition of either a holographic code or state. This is why we tentatively denote the holographic model we derive as a \textit{holographic network}. We ultimately interpret the network as the \emph{propagator} from the past boundary of de Sitter spacetime to its future boundary.

\subsection{A tessellation for de Sitter spacetime}
\label{sec:tessellation}

As a first step toward defining a holographic tensor network for $(1+1)$-dimensional de Sitter spacetime we need to introduce a \emph{tessellation}. The kind of tessellation we describe here was proposed by Aicardi \cite{aicardi_symmetries_2007}, and is comprised of causal diamonds $C(p,q)$. The endpoints of the two fundamental causal diamonds defining the tessellation lie within the temporal infinities $\mathcal{I}^+$ and $\mathcal{I}^-$.

\begin{figure}
	\begin{center}
		\includegraphics[page=4]{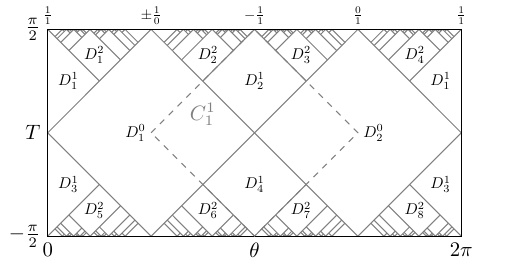}
	\end{center}
	\caption{Recursive construction of a tessellation of dS$_2$ with causal diamonds using Farey numbers.}
	\label{fig:tessellation}
\end{figure}

The tessellation, shown in Fig~\ref{fig:tessellation}, is described recursively, starting with two distinguished tiles defining the \emph{fundamental regions}. Note that, in contrast to regular tessellations of, e.g., Euclidean space, the resulting tessellation is \emph{not} invariant under a discrete subgroup of the spacetime isometry group. Further, the fundamental regions are distinguished in that they are the only tiles that extend from positive to negative infinity.

\begin{figure}
	\begin{center}
		\includegraphics[page=8]{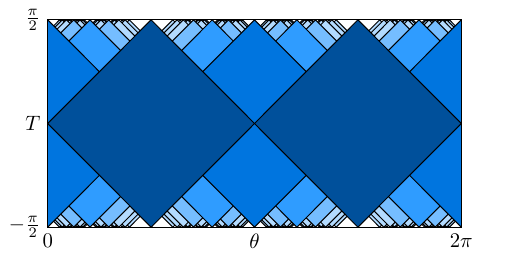}
		\caption{Tessellation of dS$_2$ (color indicates which generation the tile belongs to)}
		\label{fig:tesselRes}
	\end{center}
\end{figure}

 The causal diamonds comprising the tessellation have either one or both of their endpoints lying within the temporal boundaries. The set of all these endpoints, the \emph{boundary} of the tessellation, may be identified with the rational numbers $\mathbb{Q}$ according to the following prescription. Firstly, the two fundamental tiles, denoted $D_1^0$ and $D_2^0$ (as depicted in Fig.~\ref{fig:tessellation}), are defined to be the following causal diamonds 
\begin{align}	
	\begin{split}
		D_1^0 &= C\left(\left(\frac{\pi}{2},-\frac{\pi}{2}\right),\left(\frac{\pi}{2},\frac{\pi}{2}\right)\right) \quad \text{and}\\
		D_2^0 &= C\left(\left(\frac{3\pi}{2},-\frac{\pi}{2}\right),\left(\frac{3\pi}{2},\frac{\pi}{2}\right)\right).
	\end{split}\label{eq:fundRegions}
\end{align}
The endpoints $\pi/2$ and $3\pi/2$ (as elements of the circle $S^1\subset \mathbb{C}$) of these diamonds in $\mathcal{I}^{+}$ are identified (for reasons that will become clear) with the rational numbers $1/0$ and $0/1$ via the \emph{Cayley transformation}
\begin{equation}
	w(z) \equiv i\frac{ z-i}{z+i}.
\end{equation}
Note that in this way we have identified the positive and negative infinity of $\mathbb{Q}$ so that $1/0$ is identified with $-1/0$.
To define the next \emph{generation} of tiles we introduce the following \emph{Farey mediant operation} 
\begin{equation}
	\frac{p}{q}\oplus\frac{r}{s}=\frac{p+r}{q+s};
\end{equation}
we take the preexisting boundary points $0/1$ and $\pm 1/0$ and build, via the Farey mediant, the new points
\begin{equation}
	-\frac{1}{0}\oplus \frac{0}{1} = -\frac{1}{1} \quad\text{and}\quad \frac{0}{1}\oplus \frac{1}{0} = \frac{1}{1}.
\end{equation}
These rational numbers then induce via the Cayley transformation two new boundary points for the new causal diamonds of the first generation:
\begin{align}
	C_1^1 &= C\left(\left(\arg(w(1)),-\frac{\pi}{2}\right),\left(\arg(w(1)),\frac{\pi}{2}\right)\right) \nonumber\\&= C\left(\left(\frac{3\pi}{2},-\frac{\pi}{2}\right),\left(\frac{3\pi}{2},\frac{\pi}{2}\right)\right) \\
	C_2^1 &= C\left(\left(\arg(w(-1)),-\frac{\pi}{2}\right),\left(\arg(w(-1)),\frac{\pi}{2}\right)\right) \nonumber\\&= C\left(\left(\frac{\pi}{2},-\frac{\pi}{2}\right),\left(\frac{\pi}{2},\frac{\pi}{2}\right)\right) 
\end{align}
The tiles for the first generation are defined as the set difference from the causal diamonds defined above and the tiles of previous generations:
\begin{align}
D^1 &= \left(C_1^1 \cup C_2^1\right) / \left(D_1^0 \cup D_2^0\right).
\end{align}
This procedure continues iteratively to define tiles of higher generations:
\begin{center}
\begin{tikzpicture}[scale=8.3]
	\pgfmathsetmacro{\ya}{0.32};
	\pgfmathsetmacro{\yb}{0.24};
	\pgfmathsetmacro{\yc}{0.16};
	\pgfmathsetmacro{\yd}{0.08};
	\node[] at (0,\ya) {$\text{-}\frac10$};
	\node[] at (0.5,\ya) {$\frac01$};
	\node[] at (1,\ya) {$\frac10$};
	\foreach \y in {\yb,\yc,\yd} {
		\node[gray] at (0,\y) {$\text{-}\frac10$};
		\node[gray] at (0.5,\y) {$\frac01$};
		\node[gray] at (1,\y) {$\frac10$};}
	\node at (0.25,\yb) {$\text{-}\frac11$};
	\node at (0.75,\yb) {$\frac11$};
	\foreach \y in {\yc,\yd} {
		\node[gray] at (0.25,\y) {$\text{-}\frac11$};
		\node[gray] at (0.75,\y) {$\frac11$};}
	\node at (0.125,\yc) {$\text{-}\frac21$};
	\node at (0.375,\yc) {$\text{-}\frac12$};
	\node at (0.625,\yc) {$\frac12$};
	\node at (0.875,\yc) {$\frac21$};
	\node[gray] at (0.125,\yd) {$\text{-}\frac{2}{1}$};
	\node[gray] at (0.375,\yd) {$\text{-}\frac{1}{2}$};
	\node[gray] at (0.625,\yd) {$\frac12$};
	\node[gray] at (0.875,\yd) {$\frac21$};
	\node at (0.0625,\yd) {$\text{-}\frac{3}{1}$};	
	\node at (0.1875,\yd) {$\text{-}\frac{3}{2}$};
	\node at (0.3125,\yd) {$\text{-}\frac{2}{3}$};
	\node at (0.4375,\yd) {$\text{-}\frac{1}{3}$};
	\node at (0.5625,\yd) {$\frac13$};	
	\node at (0.6875,\yd) {$\frac23$};
	\node at (0.8125,\yd) {$\frac32$};
	\node at (0.9375,\yd) {$\frac31$};
\end{tikzpicture}
\end{center}
The resulting sequences of \emph{completely reduced} rational numbers $\frac{p}{q}$ and $\frac{r}{s}$ satisfying $\vert p s - q r\vert =1$ are called \emph{Farey numbers}. Each Farey number is then mapped to the circle via the Cayley transformation to yield the future boundary points of the subsequent generations. The past boundary points have the same $\theta$ value which is found by mirroring through the $T=0$ time slice. The resulting \emph{Farey tessellation} is illustrated in Fig.~\ref{fig:tesselRes}. The Farey mediant operation, applied recursively, exhausts the rational numbers $\mathbb{Q}$ so that the future and past boundaries of the tessellation are $\mathbb{Q}$.

The main reason for choosing the Farey tessellation is that it admits a natural action of the infinite group $\PSL(2,\mathbb{Z})$: it preserves the relation $|ad-bc|=1$, and hence mediancy. We later identify $\PSL(2,\mathbb{Z})$ as the group of isometries compatible with our tessellation (in a sense to be described).  

It turns out to be convenient to identify the Farey tessellation of de Sitter spacetime with a tessellation based on dyadic rational numbers $a/2^n$, $a\in \mathbb{Z}$, $n\in \mathbb{Z}_+$. This is achieved by applying the \emph{Minkowski question mark function $?(x)$} --- a homeomorphism of $S^1$ --- to the boundary points of the tessellation. The question-mark function may be defined recursively on the unit interval. The base cases are given by
\begin{equation}
	?(0) = 0 = \frac01 \quad\text{and}\quad ?(1) = 1 =\frac11,
\end{equation}
which are then extended recursively to all of the Farey numbers $\frac{p}{q}$ and $\frac{r}{s}$ satisfying $\vert p s - q r\vert =1$ via the rule (see, e.g., \cite{viader_new_1998} for further details):
\begin{equation}
	?\left(\frac{p}{q}\oplus\frac{r}{s}\right)  = \frac12 \, ?\left(\frac{p}{q}\right) + \frac12 \, ?\left(\frac{r}{s}\right),
\end{equation}
In this fashion one generates a correspondence between the dyadic rationals and the rational numbers in the unit interval:
\begin{center}
	\begin{tikzpicture}[scale=8]
		\pgfmathsetmacro{\ya}{0.32};
		\pgfmathsetmacro{\yb}{0.24};
		\pgfmathsetmacro{\yc}{0.16};
		\pgfmathsetmacro{\yd}{0.08};
		\node at (0,\ya) {$0$};
		\node at (1,\ya) {$1$};
		\foreach \y in {\yb,\yc,\yd} {
			\node[gray] at (0,\y) {$0$};
			\node[gray] at (1,\y) {$1$};}
		\node at (0.5,\yb) {$?(\frac12)=\frac12$};
		\foreach \y in {\yc,\yd} {
			\node[gray] at (0.5,\y) {$\frac12$};}
		\node at (0.25,\yc) {$?(\frac13)=\frac14$};
		\node at (0.75,\yc) {$?(\frac23)=\frac34$};
		\node[gray] at (0.25,\yd) {$\frac14$};
		\node[gray] at (0.75,\yd) {$\frac34$};
		\node at (0.125,\yd) {$?(\frac14)=\frac18$};	
		\node at (0.375,\yd) {$?(\frac25)=\frac38$};
		\node at (0.625,\yd) {$?(\frac35)=\frac58$};
		\node at (0.875,\yd) {$?(\frac34)=\frac78$};
	\end{tikzpicture}
\end{center}
The Minkowski question mark function may be extended to $\mathbb{Q}$ via $?(x+1) \equiv ?(x)+1$.
As a result, the function $(?(x)-x)$ is a periodic function with zeroes at the integers.

To get a tessellation that is defined on the dyadic rational numbers, the Minkowski question mark function is applied to the Farey numbers that define the boundaries of the tiles in the Farey tessellation. To define a tessellation based on dyadic rationals we need to identify the dyadic rationals on the real axis $\mathbb{R}$ with the dyadic rationals the unit interval, which is naturally identified with the unit circle. 
 
\begin{figure}
		\begin{center}
			\includegraphics[page=19]{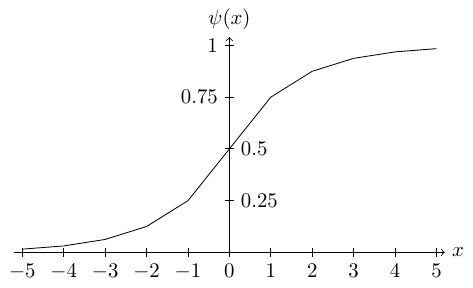}
		\end{center}
		\caption{Function $\psi(x)$ maps points from the real axis to the unit interval.}
		\label{fig:psi}
\end{figure}

 This is achieved with a particular function $\psi$ which is defined as follows. The images of subsequent integers are given by:\\
	\begin{align}
		\psi: \mathbb{Z}\rightarrow [0,1], \,\,\,\begin{cases}
			n \mapsto \dfrac{1}{2^{\vert n\vert+1}}, & n < 0\\[2ex]
			n \mapsto \dfrac12, & n = 0\\[2ex]
			n \mapsto  \dfrac{2^{ n+1}-1}{2^{ n+1}}, & n > 0,\\
		\end{cases}\label{eq:fktPsi}
	\end{align}
	which may be visualised as
	\begin{center}
		\includegraphics[page=18]{figures.pdf}
	\end{center}
	The piecewise-linear function $\psi$ is then fully specified by requiring its graph interpolates linearly between the integers. The resulting function $\psi(x)$ is plotted in Fig.~\ref{fig:psi}. 
	
Thus the dyadic rationals induce a tessellation of dS with the following iterative description. The boundary points of the fundamental tiles on the unit interval are $0$, $1/2$ and $1$. We identify these with the dyadic tessellation defined on the circle as follows:
\begin{align}	
	\begin{split}
		D_1^0 &= C\left(\left(\frac{\pi}{2},-\frac{\pi}{2}\right),\left(\frac{\pi}{2},\frac{\pi}{2}\right)\right) \quad \text{and}\\
		D_2^0 &= C\left(\left(\frac{3\pi}{2},-\frac{\pi}{2}\right),\left(\frac{3\pi}{2},\frac{\pi}{2}\right)\right).
	\end{split}\label{eq:fundRegions}
\end{align}

The first generation of tiles are defined by the region 
\begin{align}
	D^1 \equiv \left(C_1^1 \cup C_2^1 \right)  \backslash \left( D_1^0 \cup D_2^0 \right) = D_1^1 \cup D_2^1 \cup D_3^1 \cup D_4^1,
\end{align}
where
\begin{align}	
	\begin{split}
		C_1^1 &= C\left(\left(0,-\frac{\pi}{2}\right),\left(0,\frac{\pi}{2}\right)\right)\quad \text{and}
		\\
		C_2^1 &=C\left(\left(\pi,-\frac{\pi}{2}\right),\left(\pi,\frac{\pi}{2}\right)\right).
	\end{split}	
\end{align}
Accordingly, the first generation of the tessellation (i.e., the set $=D^1$) is comprised of four tiles. Each tile $D^1_i$ may be interpreted as a new causal diamond.
This procedure may be repeated recursively, giving rise to the tiles $D^n_j$ of the $n$th generation 
\begin{align}
	D^n \equiv & \left(\bigcup_{j=1}^{2^n} C_j^n\right) \setminus \left(\bigcup_{k=1}^{n-1} D^{k}\right),\\
	D^n = & \bigcup_{j=1}^{2^{n+1}} D^n_j,
\end{align}
where 
\begin{equation}
	C_j^n \equiv C\left(\left(\frac{\pi (j-1)}{2^{n-1}},-\frac{\pi}{2}\right),\left(\frac{\pi (j-1)}{2^{n-1}},\frac{\pi}{2}\right)\right).
\end{equation}

The points $p$ and $q$ defining the causal-diamond tiles $D_j^n\equiv C(p,q)$ of the dyadic tessellation lie on constant-time slices determined by the generation (note that this is not the case for the Farey tessellation): 
\begin{equation}
	T_n = \begin{cases}
		\dfrac{\pi}{2} \,\dfrac{2^n-1}{2^n}  	,\hspace{4mm} n\geq 0\\[1.5ex]
		-\dfrac{\pi}{2} \,\dfrac{2^n-1}{2^n}	,\hspace{4mm} n< 0.
	\end{cases}
	\label{eq:timeslices}
\end{equation}

\begin{figure}
	\begin{center}
		\begin{subfigure}{.48\textwidth}
			\begin{center}
				\includegraphics[page=6]{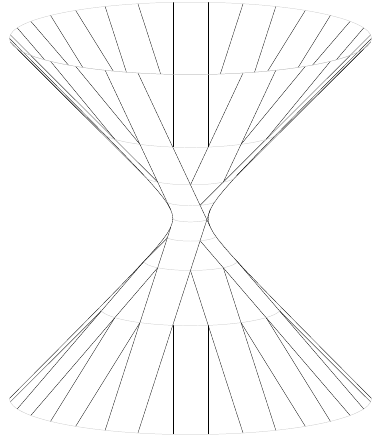}
			\end{center}
			\caption{View of the dyadic tessellation from the side.}
		\end{subfigure}\hspace{.001\textwidth}
		\begin{subfigure}{.48\textwidth}
			\begin{center}
				\includegraphics[page=7]{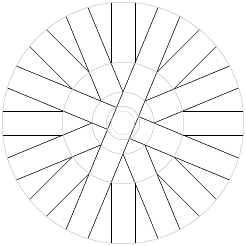}\end{center}
			\caption{View of the dyadic tessellation from the top.}
		\end{subfigure}
		\begin{subfigure}{.48\textwidth}
			\begin{center}
				\includegraphics[page=9]{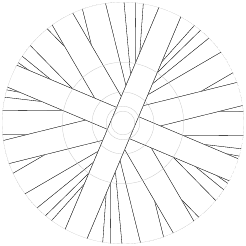}\end{center}
			\caption{View of the Farey tessellation from the top.}
		\end{subfigure}
	\end{center}
	\caption{Tessellations of dS$_2$ on the hyperbolic sheet embedded in Minkowski spacetime.}
	\label{fig:dS_2_tessel}
\end{figure}

The time slices $T_n$ admit an interesting interpretation: in the limit $T\rightarrow \infty$ the size of de Sitter spacetime doubles for each consecutive time slice. This may be observed as follows. The size 
of de Sitter spacetime at the time $\tau$ in global coordinates is given by
\begin{align}d_\text{dS}(\tau)= 2\pi  \cosh\tau.\label{eq:dSsize}
\end{align} 
The ratio of the size $d_\text{dS}(\tau_n)$ \footnote{Here $\tau_n$ denotes the time coordinate in global coordinates corresponding to $T_n$.} of de Sitter spacetime at the consecutive time slices $\tau_n$ and $\tau_{n+1}$ is given by 
\begin{equation*}
	\frac{d_\text{dS}(\tau_{n+1})}{d_\text{dS}(\tau_n)} 
	=\frac{\cos\left(\frac{\pi}{2} (1 - 2^{-n})\right)}{\cos\left(\frac{\pi}{2} (1 - 2^{-(n+1)})\right)}
	\overset{n\rightarrow\infty}{\longrightarrow}  2.
\end{equation*}
For each consecutive generation in the limit $T\rightarrow \infty$ both the number of tiles and the size of the spacetime doubles. In this way the length of the tiles in the tessellation close to the boundary approaches a constant. This is most clearly observed when one embeds de Sitter in Minkowski spacetime, see Fig.~\ref{fig:dS_2_tessel}. 

\subsection{Construction of a holographic network}
\label{sec:TNconstruct1}

In this section a holographic tensor network corresponding to $(1+1)$-dimensional de Sitter spacetime with de Sitter radius $\ell=1$ is constructed. For a review of the basics of tensor networks see Appendix \ref{app:TNbasics}. 

In order to build our tensor network we consider atoms comprised of \emph{isometries} and \emph{unitaries} (of which \emph{perfect tensors} are a subset \cite{pastawski_holographic_2015}). While the motivation for using perfect tensors comes from the AdS setting where (planar) perfect tensors ensure that the microscopic quantum tensor network inherits a residual rotation invariance, it turns out that for most of our observations we won't actually need that our tensors obey all of the conditions required of perfect tensor. Indeed, only the condition that they are isometries in the usual sense is necessary for a majority of our conclusions.

Two particular tensors are relevant for us: a 3-leg tensor $V_\alpha^{\beta\gamma}$ and a 4-leg perfect tensor $U_{\alpha\beta}^{\gamma\delta}$: 
\begin{center}
	\begin{tikzpicture}[scale=0.7]
		\node[left] at (-1.5,0) {$V^{\beta\gamma}_\alpha \equiv $};
		\draw (0,0) -- (-1.,1);
		\draw (0,0) -- (1.,1);
		\draw (0,0) -- (0,-1.5);
		\draw[fill=white] (0,0) circle (0.5);
		\node at (0,0) {$V$};
		\node[right] at (0,-1.5) {\scriptsize $\alpha$};
		\node[left] at (-1.,1) {\scriptsize $\beta$};
		\node[right] at (1.,1) {\scriptsize $\gamma$};
		\node[left] at (2,0) {$=$};
	\end{tikzpicture}
	\begin{tikzpicture}[scale=0.7]
		\draw (0,0) -- (-1.,1);
		\draw (0,0) -- (1.,1);
		\draw (0,0) -- (0,-1.5);
		\node[right,white] at (0,-1.5) {\scriptsize $\alpha$};
		\node[left,white] at (-1.,1) {\scriptsize $\beta$};
		\filldraw (0,0) circle (1mm);
	\end{tikzpicture}
\end{center}
\begin{center}
	\begin{tikzpicture}[scale=0.7]
		\node[left] at (-1.5,0) {$U^{\gamma\delta}_{\alpha\beta} \equiv $};
		\draw (0,0) -- (-1,1);
		\draw (0,0) -- (1,1);
		\draw (0,0) -- (-1,-1);
		\draw (0,0) -- (1,-1);
		\draw[fill=white] (0,0) circle (0.5);
		\node at (0,0) {$U$};
		\node[left] at (2,0) {$=$};
		\node[left] at (-1,-1) {\scriptsize $\alpha$};
		\node[right] at (1,-1) {\scriptsize $\beta$};
		\node[left] at (-1,1) {\scriptsize $\gamma$};
		\node[right] at (1,1) {\scriptsize $\delta$};
	\end{tikzpicture}
	\begin{tikzpicture}[scale=0.7]
		\draw (0,0) -- (-1,1);
		\draw (0,0) -- (1,1);
		\draw (0,0) -- (-1,-1);
		\draw (0,0) -- (1,-1);
		\node[right,white] at (1,-1) {\scriptsize $\beta$};
		\node[left,white] at (-1,1) {\scriptsize $\gamma$};
		\filldraw (0,0) circle (1mm);
	\end{tikzpicture}
\end{center}
We assume, for simplicity, that the indices of both $U$ and $V$ run over a set of size $D$. Thus $U$ and $V$ are the following two maps:
\begin{equation}\label{eq:uandvtensors}
	U:\mathbb{C}^D\otimes\mathbb{C}^D\rightarrow \mathbb{C}^D\otimes \mathbb{C}^D,\quad \text{and}\quad V:\mathbb{C}^D\rightarrow \mathbb{C}^D\otimes \mathbb{C}^D.
\end{equation}
This assumption entails no particular loss of generality (with a little work one may allow for the dimension $D$ of the legs to vary from one spacetime location to another). 
The 4-leg tensor $U$ is a unitary transformation:
\begin{align}\begin{split}
		\begin{tikzpicture}[scale=0.8]
			\node[left] at (-0.3,0) {$U^\dagger U \, =$};
			\draw (0,1.5) -- (1.5,0);
			\draw (0,0) -- (1.5,1.5);
			\draw (0,-1.5) -- (1.5,0);
			\draw (0,0) -- (1.5,-1.5);
			\draw[fill=white] (0.75,0.75) circle (0.33);
			\node at (0.75,0.75) {$U^\dagger$};
			\draw[fill=white] (0.75,-0.75) circle (0.33);
			\node at (0.75,-0.75) {$U$};
			\node[right] at (1.8,0) {$=$};
			\foreach \x in {2.8,3.8} \draw (\x,1.5) -- (\x,-1.5);
			\node[right] at (4.1,0) {$=  \,\mathbb{I}.$};
	\end{tikzpicture} \end{split}\label{tikz:U}
\end{align}
The isometric tensor $V$ is required to fulfil 
\begin{align}\begin{split}
		\begin{tikzpicture}[scale=0.8]
			\node[left] at (-1.2, 0) {$V^\dagger V \,= $};
			\draw (-1,0.3) -- (-0.5,0.7);
			\draw (0,0.3) -- (-0.5,0.7);
			\draw (-0.5,0.7) -- (-0.5,1.2);
			\draw[fill=white] (-0.5,0.7) circle (0.33);
			\node at (-0.5,0.7) {$V^\dagger$};
			\draw (-1,-0.3) -- (-0.5,-0.7);
			\draw (0,-0.3) -- (-0.5,-0.7);
			\draw (-0.5,-0.7) -- (-0.5,-1.2);
			\draw[fill=white] (-0.5,-0.7) circle (0.33);
			\node at (-0.5,-0.7) {$V$};
			\foreach \x in {0,-1} \draw (\x,0.3) -- (\x,-0.3);
			\node[right] at (0.2, 0) {$=\vphantom{P^\dagger}$};
			\draw (1.3,1.2) -- (1.3,-1.2);	
			\node[right] at (1.7, 0) {$= \, \mathbb{I}. $};
	\end{tikzpicture}\end{split}\label{tikz:V}
\end{align}
In the following the tensor-network labels $U$ and $V$, as well as the index labels for the legs, are suppressed.

\begin{figure}
	\begin{center}
		\includegraphics[page=10]{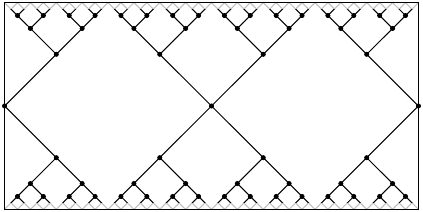}
	\end{center}
	\caption{Holographic tensor network $M$ corresponding to the tessellation Fig.~\ref{fig:tesselRes} of dS$_2$ (with $\ell=1$). This network should be understood as an infinite network extending between the temporal boundaries.}
	\label{fig:TesselTN}
\end{figure}

To construct our holographic tensor network $M$ for $\ell = 1$ de Sitter spacetime we associate tensors with the corners of the tiles as depicted in Fig.~\ref{fig:TesselTN}. The tensor network structure is obtained by contracting the legs of tensors corresponding to \emph{consecutive generations} along the edges of the tiles. The resulting network is depicted in Fig.~\ref{fig:TesselTN}. Note that the type of tensor placed at the corner of a tile corresponds to the number of neighbouring tensors. In this way we have associated the 4-leg tensor with only the initial generation. (Later we consider the construction of holographic tensor networks corresponding to dS spacetimes with radius $\ell > 1$, which feature a greater number of 4-leg tensors.) The only uncontracted legs of the resulting \emph{infinite} tensor network are associated with the temporal boundaries of the tensor network. 

\subsection{Kinematic Hilbert space}\label{subsubsec:kinhilbertspace}
One of the most important features of de Sitter spacetime is that all the observables live on the temporal boundaries $\mathcal{I}^{\pm}$. As a consequence, configurations --- kinematical states --- which determine the expectation values of the observables are associated with the temporal boundaries. This simple observation has profound consequences for the structure of the kinematical Hilbert space. Indeed, we see little alternative except to associate the Hilbert space of states of dS with the boundaries $\mathcal{I}^{\pm}$ and hence introduce two infinite-dimensional Hilbert spaces $\mathcal{H}_{\text{in}}$, the \emph{input} space, and $\mathcal{H}_{\text{out}}$, the \emph{output} space, associated with these boundaries, respectively.

Our focus in this paper is on the infinite tensor network $M$ in Fig.~\ref{fig:TesselTN} to be understood as an operator from the (as yet to be defined) infinite-dimensional Hilbert space $\mathcal{H}_{\text{in}}$ to $\mathcal{H}_{\text{out}}$. There are considerable technicalities that must be overcome to discuss these Hilbert spaces at the level of mathematical rigour. We do not undertake this investigation here, and the arguments in this paper are at a physical level of rigour. (The interested reader is directed to \cite{Osborne2017_DynHolCode} and \cite{osborne_quantum_2019} where the construction of these Hilbert spaces as inductive limits are described in detail.) 

Instead of introducing the formalism of inductive limits, we equivalently reason about our infinite tensor-product Hilbert spaces, and the network $M$, by appealing to cutoff versions of the network: we frequently draw only a couple of generations of the network and use the isometric and unitary property of $U$ and $V$ to deduce algebraic facts about the infinite network $M$, a strategy successful because these conditions can lead to an infinite number of cancellations, particularly in equations involving $M$ and its adjoint $M^\dag$. 

We describe the cutoff Hilbert spaces corresponding to $\mathcal{H}_{\text{in}}$ and $\mathcal{H}_{\text{out}}$ as follows. Choose a nonnegative integer $n\ge 0$: we begin by discretising the spatial coordinate $\theta$ by breaking the circle into $2^n$ \emph{dyadic} intervals $2\pi [j 2^{-n}, (j+1)2^{-n})$, where $j \in \{0,1,\ldots, 2^n-1\}$. These discretisations are associated, respectively, with the time slices $T_{\pm n}$. Attached to each such time slice $T_n$ is the finite Hilbert space
\begin{equation}
	\mathcal{H}_{\pm n} \equiv \bigotimes_{j=0}^{2^n-1} \mathbb{C}^D.
\end{equation}
The tensor network defined by our tessellation induces, for all $m\le n$, a \emph{family} of linear maps 
\begin{equation}
	M_{m,n}:\mathcal{H}_m\rightarrow \mathcal{H}_n,
\end{equation}
given by drawing the tensor network for the appropriate number of generations and taking the dangling legs to act on the respective input and output spaces $\mathcal{H}_m$ and $\mathcal{H}_n$. Note that, for $1 < m < n$ the operator $M_{m,n}$ is an isometry obeying 
\begin{equation}
	M_{m,n}^\dag M_{m,n} = \mathbb{I}_{\mathcal{H}_m}.
\end{equation}
Conversely, for $m<n<-1$, the operator $M_{m,n}$ is an isometry satisfying
\begin{equation}
	M_{m,n} M_{m,n}^\dag = \mathbb{I}_{\mathcal{H}_n}.
\end{equation}

The tensor network $M$ then acts ``in the limit'' $m\rightarrow -\infty$ and $n\rightarrow \infty$. To make proper mathematical sense of this statement one needs the technology of inductive limits. However, this is equivalent to the expedient of always working in the largest input and output Hilbert spaces (corresponding to the finest required discretisations) needed to express all the physical statements required. With a little care the Hilbert spaces $\mathcal{H}_{\text{in}} \text{``$\equiv$''} \mathcal{H}_{-\infty}$ and $\mathcal{H}_{\text{out}} \text{``$\equiv$''} \mathcal{H}_{\infty}$ may be simply regarded as the infinite tensor product of $\mathbb{C}^D$ \footnote{Care must be exercised here: the infinite tensor product Hilbert space should be regarded as the kinematical space built from a given reference state $|\Omega\rangle$ and all states built from it by applying local operations.}.

The actual definition of $\mathcal{H}_{\text{in}}$ and $\mathcal{H}_{\text{out}}$ is intimately tied to the tensor network $M$ as we use the family $M_{m,n}$ to build an \emph{equivalence relation} on the cutoff Hilbert spaces $\mathcal{H}_{m}$. Suppose that $1<m<n$: we say that $|\phi_m\rangle \in \mathcal{H}_{m}$ and $|\psi_{n}\rangle \in \mathcal{H}_{n}$ are \emph{equivalent}, written $|\phi_m\rangle\sim |\psi_{n}\rangle$, if
\begin{equation}
 	|\psi_n\rangle = M_{m,n}|\phi_m\rangle.
\end{equation} 
Physically we think of  $|\psi_n\rangle$ as a \emph{fine graining} of $|\phi_m\rangle$.
(The analogous definition holds for the case $m<n<-1$.) In this way we realise the cutoff Hilbert space $\mathcal{H}_m$ as a subspace of the fine-grained Hilbert space $\mathcal{H}_n$. The definition of $\mathcal{H}_{\text{out}}$ now results by requiring it to be the smallest Hilbert space which contains all of $\mathcal{H}_m$, $m > 2$ as subspaces, via $M_{m,\infty}$.

\subsection{Basic properties of the tensor network}
De Sitter spacetime is characterised by an initial \emph{contraction} for $T<0$, followed by \emph{expansion} for $T>0$. Implementing these two epochs at a microscopic quantum level immediately throws up conceptual challenges. At a naive level a contracting spacetime should entail the \emph{loss}, \emph{destruction}, or \emph{deletion} of quantum information.  As the spacetime contracts there is simply less volume and hence less information may be stored in the quantum degrees of freedom comprising it. Conversely, for $T>0$, dS is expanding and new quantum degrees of freedom are brought into existence. There are many ways to model loss/deletion and creation processes in quantum mechanics. The simplest strategy, and the one employed here, is to model such processes with \emph{isometries}. (Another possibility, which doesn't violate unitary so directly, is to model a loss process with a \emph{completely positive map}. This, however, does introduce a new mystery: where does the information go?) 

To make this discussion more concrete we consider the Hilbert space $\mathcal{H}_{\text{in}}$ for the degrees of freedom on the temporal boundary $\mathcal{I}^-$, $\mathcal{H}_0$ for the degrees of freedom at $T=0$, and $\mathcal{H}_{\text{out}}$ for $\mathcal{I}^+$. According to the above argument the contraction process should be represented at a microscopic level by an isometry $A^\dag:\mathcal{H}_{\text{in}}\rightarrow \mathcal{H}_0$ and the expansion process by an isometry $B:\mathcal{H}_0\rightarrow \mathcal{H}_{\text{out}}$. Hence the entire spacetime history should be represented by the operator
\begin{equation}
	W=BA^\dag : \mathcal{H}_{\text{in}}\rightarrow \mathcal{H}_{\text{out}}
\end{equation}
which can be thought of as a scattering process. Operators $W$ which are such compositions of isometries are known as \emph{partial isometries}: recall that a bounded linear transformation $C:\mathcal{H}_1\rightarrow\mathcal{H}_2$ is called a partial isometry, if $P=C^\dagger C$ is a projection. Choosing $C = W$ we deduce that $W^\dag W = A B^\dag B A^\dag = A A^\dag$ is a projection (similarly, so is $W^\dag$). In this way we intepret the operator $W$ as a kind of \emph{restricted propagator} from the Hilbert space of the temporal past infinity to the Hilbert space of the temporal future infinity. In general the operator $W$ fails to be a propagator in the usual sense because $W^\dag W$ is only a projection rather than the identity required for $W$ to be unitary. 

The projection $P=W^\dag W$ arising from a partial isometry $W:\mathcal{H}_{\text{in}}\rightarrow \mathcal{H}_{\text{out}}$ singles out a distinguished subspace of $\mathcal{H}_{\text{in}}$ via the projection $P$, i.e., $\mathcal{H}_{\text{phys}}\equiv P\mathcal{H}_{\text{in}}$, which we term the subspace of \emph{physical states}. A physical state $|\phi_{\text{phys}}\rangle \in \mathcal{H}_{\text{phys}}$ propagates without loss of norm through the spacetime network.

Consider the tensor network $M$ depicted in Fig.~\ref{fig:TesselTN}: to check whether $M^\dagger M$ is a projection we exploit the identities (\ref{tikz:U}) and (\ref{tikz:V}):\\

\begin{center}\includegraphics[page=29]{figures.pdf}\end{center}
Hence the tensor network $M$ is indeed a partial isometry as $M^\dagger M$ is a projection:\\
\begin{center}\includegraphics[page=30]{figures.pdf}\end{center}

A key property of the tensor network $M$ representing the evolution from $\mathcal{H}_{\text{in}}$ to $\mathcal{H}_{\text{out}}$ is that the subspace $\mathcal{H}_{\text{phys}}\equiv (M^\dag M) \mathcal{H}_{\text{in}}$ of physical states determined by $M$ is \emph{finite dimensional}. If the dimension of the legs of $U$ and $V$ is $D$ we have that $M^\dag M$ is a projection onto a $D^4$-dimensional subspace of $\mathcal{H}_{\text{in}}$. We later connect this observation with the $\Lambda$ - $N$ correspondence.  

The tensor network $M$ exhibits an \emph{information bottleneck} at $T=0$: the infinite-dimensional input Hilbert space $\mathcal{H}_{\text{in}}$ has to pass through the finite-dimensional $\mathcal{H}_{\text{phys}}$ subspace. It is notable that this phenomena whereby a finite-dimensional subspace canonically emerges from an infinite-dimensional ambient Hilbert space is most transparently observed in the tensor-network representation.  

The discussion in this section mirrors that of Witten \cite{witten_quantum_2001} concerning the nonperturbative definition of the Hilbert space for dS. Indeed, we are motivated to conjecture that the subspace $\mathcal{H}_{\text{phys}}$ introduced here is a microscopic realisation of Witten's nonperturbative Hilbert space $\mathcal{H}$. Further, the partial isometry $W$ is a microscopic realisation of the matrix $M$ constructed by Witten. 

So far our discussions have centred around a tensor network representation for $\ell = 1$. The question of how to represent dS spacetimes with larger radii and, further, the limit $\ell\rightarrow \infty$ now emerges. There is a natural construction, as we detail in the next subsection.

Before we turn to this construction we pause to highlight the general nature of the arguments in this section: the essential ingredient is that a tensor network representing dS may be written as a composition of two isometries, namely, an isometry, representing contraction, mapping from the (large) input Hilbert space representing $\mathcal{I}^-$ to a smaller intermediate Hilbert space $\mathcal{H}_0$ composed with an isometry acting from $\mathcal{H}_0$ to the final Hilbert space for $\mathcal{I}^+$. This abstract structure is not restricted to $(1+1)$ dimensions. Indeed, with little modification one may observe the same structure of tensor networks representing dS in higher dimensions.

\subsection{Holographic networks for de Sitter spacetime with $\ell\not= 1$: a $\Lambda$-qubit correspondence}
We have, so far, only considered tessellations and tensor networks corresponding to a $(1+1)$-dimensional de Sitter spacetime with radius $\ell=1$. In this section we explain how to generalise the construction to model $(1+1)$-dimensional de Sitter spacetimes with different de Sitter radii $\ell\not=1$. According to this definition one obtains a correspondence between the network quantum information capacity of the network and the cosmological constant. This is reminiscent of Bousso's $\Lambda$-$N$ correspondence \cite{bousso_positive_2000}.

We illustrate our definition for de Sitter radii $\ell=2$ and $\ell=4$; the construction can be readily generalised to describe de Sitter spacetimes with radii $\ell=2^n$. The definition is intended to remain consistent with the causal structure of the tensor network for $\ell=1$. To achieve this we modify the tessellation by dividing the two fundamental regions into four smaller causal tiles, respectively. In this way, the tensor network is only locally modified around the time slice $T_0$: this is physically plausible as the finiteness of the de Sitter radius only obtrudes itself most profoundly around $T_0$, and its effects are progressively less significant towards the temporal infinities.
\begin{figure}
	\begin{center}
		\includegraphics[page=11]{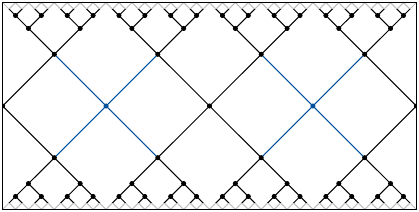}
	\end{center} 
	\caption{Tessellation and holographic network for dS$_2$ with $\ell=2$. The blue lines (edges of $D^1$) divide the initial causal diamonds.}
	\label{fig:TesselTN_l2}
\end{figure}
The construction of the modified tensor network then proceeds analogously to Sec.~\ref{sec:TNconstruct1} and is depicted in Fig.~\ref{fig:TesselTN_l2}. This tensor network is a partial isometry, just as for $\ell=1$ (see Appendix \ref{app:PIlneq1}). 

\begin{figure}
	\begin{center}
		\includegraphics[page=12]{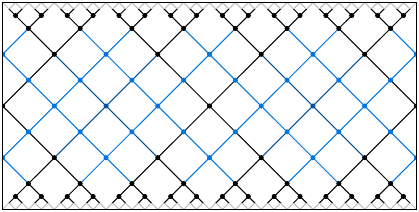}
	\end{center}
	\caption{Tessellation and holographic network for dS$_2$ with $\ell=4$. The blue (edges of $D^1$) and light blue (edges of $D^2$) lines divide the initial causal diamonds.}
	\label{fig:TesselTN_l4}
\end{figure}

The construction of a tensor network for radius $\ell=4$ and larger proceeds iteratively as above: the tiles of the first two generations, i.e., the fundamental and first, in the tessellation are each subdivided into four diamonds. Around $T=0$ this new tessellation now has four times as many tiles as the initial tessellation. Positing that each tile represents a quantum spacetime degree of freedom leads us to conclude that the new tessellation represents a spacetime four times as large as the original. The corresponding holographic tensor network is depicted in Fig.~\ref{fig:TesselTN_l4}. The resulting network is also a partial isometry.

Given our definition for $\ell>1$ we now deduce a fundamental relation connecting the cosmological constant and the quantum information capacity of the corresponding tensor network. As the network $M_{\ell}$ is a partial isometry, one obtains a corresponding completely positive (CP) -- but not trace-preserving -- map via
\begin{equation}
	\mathcal{E}_{\ell}(\rho) \equiv M_{\ell}\rho M^\dag_{\ell}.
\end{equation}
This superoperator takes density operators $\rho_{\text{in}}$ for the initial Hilbert space $\mathcal{H}_{\text{in}}$ to density operators for $\mathcal{H}_{\text{out}}$. A CP map can be understood as representing a communication process from a sender to a receiver. In this context it makes sense to ask what the corresponding \emph{quantum capacity} is of this process, i.e., how many qubits $Q_\ell^{(1)}$ of quantum information can be sent (perhaps better: \emph{stored}) in a single use, without error through the CP map. This is easily obtained in our context as 
\begin{equation}
	Q_\ell^{(1)}=\lfloor\log_2(\dim(\mathcal{H}^\ell_{\text{phys}}))\rfloor.
\end{equation}
This follows because $M_\ell$ is a partial isometry and it is possible to directly encode $2^{Q^{(1)}_\ell}$ qubits into the subspace $\mathcal{H}_{\text{phys}}$ which is then transmitted noiselessly. Note that \footnote{Recall that $D$ is the dimension of the legs of the tensors.} 
\begin{equation}\label{eq:qcapacityTN}
	Q^{(1)}_\ell \lesssim (\ell+1)\log_2(D).
\end{equation}
This quantity admits a transparent physical interpretation: it simply counts the number of qubits that can be sent without disturbance through the bottleneck of the tensor network at $T_0$.

The correspondence (\ref{eq:qcapacityTN}) implies a fundamental relationship between the \emph{cosmological constant} $\Lambda$ and the \emph{information carrying capacity} of the network representing the spacetime. In our case this relationship reads
\begin{equation}\label{eq:infocc}
	Q^{(1)} \propto \frac{1}{\sqrt{\Lambda}}.
\end{equation}
This is a direct consequence of the equation \footnote{We are unwilling, at this premature stage, to posit the constant of proportionality.}
\begin{align}
	\frac{\ell}{\sqrt{3}} = \sqrt{\frac{1}{\Lambda}}
\end{align}
for the dS radius \cite{bousso_adventures_2002}.

If we take the correspondence (\ref{eq:infocc}) at face value and extrapolate it to the logical extreme one concludes that a positive cosmological constant determines the information carrying capacity of a spacetime and, conversely, a spacetime with a finite quantum communication capacity corresponds to a spacetime with a positive cosmological constant. A most dramatic consequence of the correspondence (\ref{eq:infocc}) is that our apparently continuous and infinite universe is effectively described by a \emph{finite-dimensional} Hilbert space carrying the quantum information from past to future. (However, one should be appropriately cautious: the example of AdS does strain this conclusion somewhat, i.e., what is implied by a negative cosmological constant? We have no answer to this, except to note that the structure of the observables of AdS is radically different to $\Lambda > 0$ spacetimes.)

The discussion here is strongly reminiscent of the $\Lambda$--$N$ correspondence, introduced by Bousso \cite{bousso_positive_2000}. Here a correlation between the cosmological constant $\Lambda$ and the number $N$ of quantum degrees of freedom describing a spacetime is posited. The argument put forward by Bousso to justify the $\Lambda-N$ correspondence builds on the key observation that the total entropy perceived by an observer $\mathcal{O}$ carrying out an experiment commencing at $p$ and ending at $q$ is bounded by that of the causal diamond $C(p,q)$. 

Bousso's argument may be understood in our context as follows: since the laboratory for a local observer $\mathcal{O}$ can only execute unitary operations local to $\mathcal{O}$ then, if we assume that the observer $\mathcal{O}$ initialises their laboratory apparatus in a pure state $|\Omega\rangle$ at $p$, the total amount of entropy that $\mathcal{O}$ can \emph{create} by losing halves of entangled pairs before the experiment concludes at $q$ is bounded by the number of qubits crossing the boundary of $\mathcal{J}^{-}(q)$. The amount of entropy that $\mathcal{O}$ can \emph{receive} before the experiment is complete is bounded by the number of qubits crossing the boundary $\mathcal{J}^{+}(p)$. The sum of these entropies is thus determined by the flux of qubits through the boundary of the causal diamond. Exploiting the \emph{covariant entropy bound} \cite{bousso_covariant_1999} one can then argue that the entropy of a causal diamond is bounded by its area.

\subsection{Causal structure of the tensor network}

It may be observed that as the de Sitter radius $\ell$ is increased toward infinity, the tensor network resulting from the above procedure has the structure of an increasingly large regular grid of unitary operators around the timeslice $T_0$ preceded, and followed, by tree-like tensor networks. In the limit $\ell\rightarrow \infty$ we should, and do, recover Minkowski spacetime in the sense that the tensor network is essentially comprised of a regular grid of unitaries, a \emph{quantum cellular automaton} (QCA) \cite{schumacher2004reversible,arrighi2020quantum,arrighi2003note,bibeau2015doubly,Bisio2017,DEBBASCH2019340}.  Such networks also arise when applying the Lie-Trotter decomposition to the dynamics of quantum spin chains. Due to the natural causal structure exhibited by QCA, namely that information propagation is bounded by a speed of light, they provide a natural candidate for a tensor-network realisation of flat Minkowski spacetime. This proposal has been explored recently in a variety of settings \cite{beny2013causal,czech_tensor_2016,de2016holographic,milsted2018tensor,cotler2019quantum,kunkolienkar2017towards,bhattacharyya2018notes}.

It turns out that the tensor-network ansatz we propose here, namely a QCA preceded and followed by tree-like tensor networks, is most closely related to that of a \emph{Lorentzian MERA} introduced in \cite{milsted2018tensor,milstedGeometricInterpretationMultiscale2018a}. In these papers the Lorentzian MERA has been argued to capture the dynamics of quantum systems on $\text{dS}_2$. The emergence, in the limit $\ell\rightarrow \infty$, of a QCA structure in our tensor-network ansatz, and the connection to Lorentzian MERA, provides further supporting evidence that our tensor network provides a microscopic model for de Sitter spacetime.  

As our tensor network is comprised of unitaries and isometries it exhibits, similar to MERA, a natural causal structure. One can confirm this structure using a variety of notions of \emph{quantum causal influence}, see \cite{beny2013causal,cotler2019quantum} and references therein for a cross-section of approaches. For example, exploiting the notion of \emph{pure causality} introduced by B\'eny in \cite{beny2013causal}, one directly recovers the causal structure of de Sitter spacetime.  

\section{Symmetries of de Sitter spacetime}
In this section we discuss the symmetries of de Sitter spacetime. Ultimately our goal here is to understand the action of these symmetries on the kinematical Hilbert spaces $\mathcal{H}_{\text{in}}$ and $\mathcal{H}_{\text{out}}$ associated with our tensor network. To this end we first review the action of isometries on the temporal boundaries. 

We can infer the isometries of dS$_2$ by first noting that the metric of three dimensional Minkowski spacetime $\mathbb{R}^{1,2}$ (in which an embedding of de Sitter spacetime is found) is preserved by a transformation $O\in \oo(1,2)$ \cite{nomizu_lorentz-poincare_1982}:
\begin{equation}
	\eta = O^T \eta O . \label{eq:trafo}
\end{equation}
This transformation induces a symmetry of  $(1+1)$-dimensional de Sitter spacetime because the equation defining the embedding of dS is preserved:
\begin{equation}
	x^T \eta x = x^T 	O^T \eta O x = (O x)^T \eta O x= y^T \eta y = 1.
\end{equation}

The action of isometries $O\in \oo(1,2)$ on dS can be equivalently specified in terms of their action on null geodesics, which are specified by the parameters $u$ and $v$ (see Sec.~\ref{sec:nullgeodesics}). 
This, in turn, allows us to describe the action of isometries on the temporal boundaries. With this action in hand we are then able to describe the action of isometries on the kinematic Hilbert spaces $\mathcal{H}_{\text{in}}$ and $\mathcal{H}_{\text{out}}$. To simplify our discussion, we restrict our attention in the following to the subgroup $\SO(1,2)$ of proper isometries preserving orientation.

\subsection{Symmetry action on the future boundary of de Sitter spacetime}
\label{sec:Mobius}

The temporal boundaries of $(1+1)$-dimensional de Sitter spacetime are circles. Accordingly, it can be helpful to describe the symmetry action of an isometry $O$ on the boundary as a linear fractional transformation of the circle or \emph{M\"obius transformation}. That this is possible is a consequence of a sporadic isogeny of $\SL(2,\mathbb{R})$ to the symmetry group $\SO(1,2)$, which we detail in Appendix~\ref{app:symmMatrix}. According to this correspondence an element
\begin{align}
	g=\begin{pmatrix} a & b \\ c & d \end{pmatrix},\quad g\in \SL(2,\mathbb{R}),
\end{align}
is identified with a corresponding symmetry transformation in $h(g)\in \SO(1,2)$. The matrix elements of $h(g)$ are somewhat complicated, and we refer to Appendix~\ref{app:symmMatrix} for further details. Note that this correspondence is $2$ to $1$: the element $g' = -g$ yields the same transformation as $g$. In this way the \emph{projective special linear group}
\begin{align}
	\PSL(2,\mathbb{R}) = \SL(2,\mathbb{R}) / \{\pm \mathbb{I}\}
\end{align}
emerges as the natural subgroup to identify with the isometries of dS.  

There is an identification between elements of $\PSL(2,\mathbb{R})$ and elements of the \emph{M\"obius group} of linear fractional transformations of the upper half plane
\begin{align}\mathbb{H} = \{z\in \mathbb{C}: \real z \geq  0\}\end{align} 
given by
\begin{align}
	f: \mathbb{H} \rightarrow \mathbb{H}, \,\,\, z\mapsto \frac{\alpha z + \beta}{\gamma z + \delta}, \label{eq:MoebiusFkt}
\end{align}
where $\alpha$, $\beta$, $\gamma$ and $\delta$ are real coefficients satisfying the condition $\alpha \delta - \beta \gamma =1$. 
Happily, as we describe below, this identification is compatible with the action of $\SO(1,2)$ on the temporal boundaries $\mathcal{I}^{\pm}$.

A temporal boundary of de Sitter spacetime may be rescaled to the unit circle $S^1$, which may be understood as the boundary of the unit disk 
\begin{align}
	\mathbb{D} = \{ z \in \mathbb{C}: \vert z\vert \leq 1 \}
\end{align}
in the complex plane.
We now identify the unit circle with the upper half plane via the Cayley transform $w$ and its inverse:
\begin{align}
	w: \mathbb{H}\rightarrow \mathbb{D}, \,\,\, z \mapsto i\frac{ z-i}{z+i}, \\ 
	w^{-1}: \mathbb{D}\rightarrow \mathbb{H}, \,\,\, z\mapsto  \frac{z+i}{1+i z}.
\end{align}
The Cayley transform identifies the real axis $\mathbb{R}$ (the boundary of the upper half plane) with the unit circle $S^1$ (the boundary of the unit disk); a boundary point of de Sitter spacetime can now be identified with the complex number
\begin{align} 
	z = u + i v \quad \text{where} \quad u,v\in \mathbb{R},
\end{align}
with $u^2+v^2=1$.

We now obtain the induced action of a M\"obius transformation $f$ on the temporal boundary $\mathcal{I}^+$ according to
\begin{align}
	p = w \circ f \circ w^{-1}: \mathbb{D}\rightarrow \mathbb{H} \rightarrow \mathbb{H} \rightarrow \mathbb{D}. \label{eq:pSymm}
\end{align}
The induced image of a point $z$ in $S^1$ under $f$ is thus
\begin{align}
	z' = p(z) \quad \Rightarrow \quad \begin{array}{c}
		u' = \real p(z) \\ v' = \imag p(z)
	\end{array}.
\end{align}
An explicit computation yields:
\begin{subequations}\begin{align} 
		u' &=   \frac{4 \alpha \delta u +2\beta \delta  (1-  v) -2u+2\alpha  \gamma  (v+1)}
		{2 u (\alpha  \beta+\gamma  \delta)  +(\alpha ^2+\gamma ^2) (v+1)-(\beta^2+\delta^2) (v-1)}\\
		v' &= \frac{4 \alpha  \beta  u+2 \alpha ^2 (v+1)-2 \beta ^2 (v-1)}
		{2 u(\alpha  \beta  + \gamma  \delta  )+(\alpha^2+\gamma ^2)  (v+1)-(\beta ^2 +\delta) ^2 (v-1)}-1
	\end{align}\label{eq:uvMoebius}\end{subequations}
As we detail in Appendix~\ref{app:symmMatrix} this action exactly matches the action of $\SO(1,2)$ on the temporal boundary found by a direct computation involving null geodesics.

\begin{figure}
	\begin{center}
		\includegraphics[page=14]{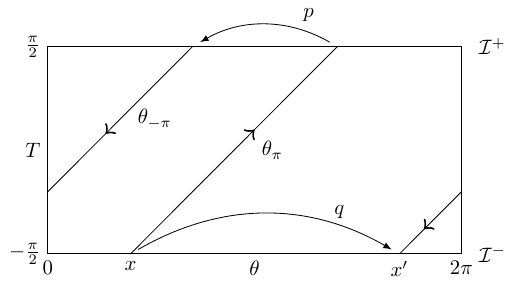}
	\end{center}
	\caption{Symmetry action $q$ on the past boundary $\mathcal{I}^-$ of dS$_2$ 
	}
	\label{fig:SymmetryI-}
\end{figure}

The action of an isometry on the future boundary $\mathcal{I}^+$ of de Sitter spacetime may be immediately extended to obtain an action on the past boundary $\mathcal{I}^-$. In order to do so, we identify a point $x$ in $\mathcal{I}^-$ with a point $x'\in\mathcal{I}^+$ by transporting it along a null geodesic. The symmetry action is applied to $x'$ and the result transported back $\mathcal{I}^-$ via a null geodesic. This action is well defined because null geodesics travelling in either direction between $\mathcal{I}^+$ and $\mathcal{I}^-$ transport a past boundary point to the same future boundary point. Explicitly, the maps $\theta_{\pm\pi}:\mathcal{I}^{\mp}\rightarrow \mathcal{I}^{\pm}$ transporting points between the temporal boundaries are given by $\theta_{\pm\pi}(x) = x \pm \pi$. Hence we obtain for an isometry $p$ on the future boundary the induced action on the past boundary $\mathcal{I}^-$, depicted in Fig.~\ref{fig:SymmetryI-}, via 
\begin{align}
	q = \theta_{-\pi} \circ p \circ \theta_\pi.
\end{align}

\subsection{Symmetry action of $\text{PSL}(2, \mathbb{Z})$ on a tessellation}
Just as general isometries of $\mathbb{R}^n$ are incompatible with a regular grid, the action, via Möbius transformations, of the spacetime isometries of dS cannot be compatible with our tessellation. This is because boundary points of the tessellation are, in general, sent to points lying outside the tessellation. We can, however, identify a discrete subgroup of $\SL(2,\mathbb{R})$ which is \emph{compatible} with the tessellation in the sense that the \emph{set of boundary points} of the tessellation is left invariant. As we'll see, the tessellation itself is not invariant; this has important consequences for the interpretation of our tensor network.       

We have seen, in the previous section, how the isometry group $\SL(2,\mathbb{R})$ acts on the temporal boundaries of dS. However, the boundary points of our tessellation (in the original non-dyadic formulation) are precisely the rational numbers lying in the unit circle $\mathbb{Q}$. Since the preimage of a rational point under the Cayley transform $w$ is rational, we see that the subgroup of isometries compatible with the tessellation is then $\PSL(2,\mathbb{Q})$. It turns out that the action of this group is actually equivalent to that of $\PSL(2,\mathbb Z)$: consider a M\"obius transformation $f$ associated with an element of $\PSL(2,\mathbb{Q})$ with arbitrary rational entries. Clearing denominators allows us to write
\begin{align}
	f(z) = \frac{\frac{a_1}{a_2} \,z + \frac{b_1}{b_2}}{\frac{c_1}{c_2} z + \frac{d_1}{d_2}} 
	= \frac{ {a_1} {b_2} {c_2} {d_2}\, z +{a_2} {b_1}{c_2} {d_2}}{	{a_2} {b_2}{c_1} {d_2} z+{a_2} {b_2}{c_2} {d_1}}.
\end{align}
Here the parameters $a_1$, $a_2$, $b_1$, $b_2$, $c_1$, $c_2$, $d_1$ and $d_2$ as well as their product are integers. Thus any transformation in $\PSL(2,\mathbb{Q})$ is equivalent to a corresponding transformation in $\PSL(2,\mathbb{Z})$.  This justifies using  $\PSL(2,\mathbb{Z})$ as the group of isometries of the Farey tessellation.

\begin{figure}
	\begin{center}
		\includegraphics[page=15]{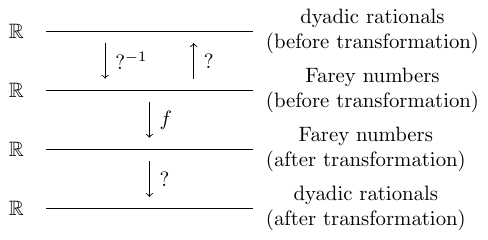}
	\end{center}
	\caption{Symmetry transformation on the dyadic rational numbers induced by a M\"obius function $f$.}
	\label{fig:transR}
\end{figure}

\begin{figure}
	\begin{subfigure}{.4\textwidth}
		\begin{center}
			\includegraphics[page=16]{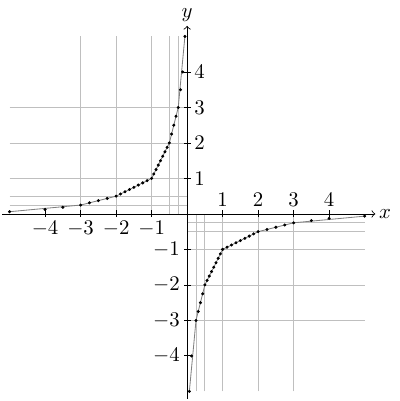}
		\end{center}
		\caption{transformation with $f(x)=-\frac1x$}\end{subfigure}
	\begin{subfigure}{.4\textwidth}
		\begin{center}
			\includegraphics[page=17]{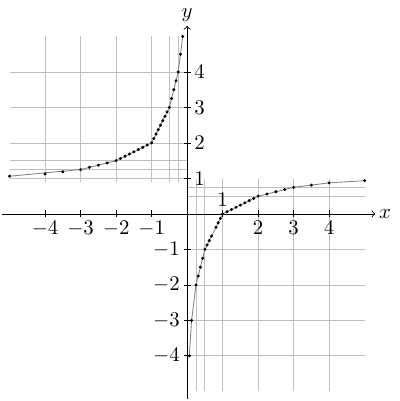}
		\end{center}
		\caption{transformation with $f(x)=\frac{x-1}{x}$}
	\end{subfigure}
	\caption{Different symmetry transformations $(?\circ f \circ ?^{-1})(x)$ with M\"obius functions $f$ which act on the dyadic rational numbers of the real axis. }
	\label{fig:?f?}
\end{figure}

Since it has been convenient to formulate our tessellation in terms of the dyadic rationals, we also describe how to obtain an induced action of $f \in \PSL(2,\mathbb{Z})$ on the subset of all dyadic rationals in $\mathbb{R}$. This is achieved by composing with the Minkowski question mark function $?$ and its inverse $?^{-1}$, i.e., the action is described via 
\begin{equation}
	(?\circ f\circ ?^{-1})(x).
\end{equation}
The steps of this symmetry transformation -- plotted in Fig.~\ref{fig:?f?} for a pair of illustrative examples -- are illustrated in Fig.~\ref{fig:transR}. The resulting symmetry transform dilates and compresses spacetime along light rays.

\begin{figure}
	\begin{subfigure}{0.445\textwidth}
		\begin{center}
			\includegraphics[page=20]{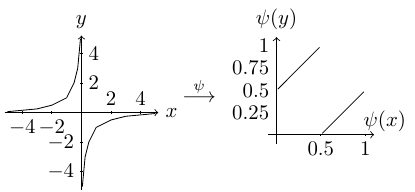}
		\end{center}
		\caption{transformation $f(x)=-\frac1x$}
	\end{subfigure}
	\begin{subfigure}{0.445\textwidth}
		\begin{center}
			\includegraphics[page=21]{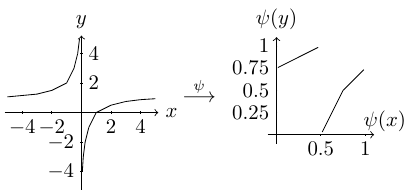}
		\end{center}
		\caption{transformation $f(x)=\frac{x-1}{x}$}
	\end{subfigure}
	\caption{Symmetry transformations $(?\circ f\circ ?^{-1})(x)$ for different M\"obius functions $f$  mapped to the unit interval using $\psi(x)$.} 
	\label{fig:PSLGenerators}
\end{figure}

The function $(?\circ f \circ ?^{-1})(x)$ is a piecewise-linear homeomorphism of $\mathbb{R}$. In order to obtain an action on the boundaries of de Sitter spacetime we further identify the domain and range of this function with the circle.  This is achieved with the function $\psi$ that was introduced earlier (see Fig. \ref{fig:psi}) and identifies the dyadic rationals on the real axis $\mathbb{R}$ with the dyadic rationals in the unit interval. In this way we finally obtain an action on the defining boundary points of our tessellation. 
Using the piecewise-linear function $\psi$, $(?\circ f\circ ?^{-1})$ can now be mapped to the unit interval. The result is plotted in Fig.~\ref{fig:PSLGenerators}.

The projective special linear group $\PSL(2,\mathbb{Z})$ may be presented as
\begin{align}
	\langle a,b \vert a^2 = b^3 = 1 \rangle \label{eq:repPSL},
\end{align}
where the generators $a$ and $b$ are $2\times 2$ matrices given by
\begin{align}
	a = \begin{pmatrix} 0 & -1 \\ 1 & 0	\end{pmatrix}
	\quad \text{and} \quad
	b = \begin{pmatrix} -1 & 1 \\ -1 & 0	\end{pmatrix}.
\end{align}
The transformations considered in Fig.~\ref{fig:?f?} and Fig.~\ref{fig:PSLGenerators} correspond precisely to $a$ and $b$ according to $f(x)=-\frac1x$ and $f(x)=\frac{x-1}{x}$. By composing these two functions (and their inverses) we hence obtain a representation of the action of $\PSL(2,\mathbb{Z})$ on the circle via piecewise-linear functions compatible with the dyadic rationals. The resulting action on the tessellation is described in the next section.

\section{Asymptotically de Sitter spacetimes}
De Sitter spacetime in $(1+1)$ dimensions arises as a solution to 2D Einsten gravity, however this theory is trivial as it reduces to the Euler characteristic. Further, it is impossible to couple to matter fields without confining to the ground state or putting up with deleterious backreation. A natural generalisation of two-dimensional gravity which avoids these difficulties  is furnished by the \emph{Jackiw-Teitelboim} (JT) model \cite{jackiw1985lower,teitelboim1983gravitation,jackiw1992gauge}. This theory allows for more possibilities and has nearly-dS solutions. Although JT gravity does not have bulk gravitons, it does have boundary gravitons arising from fluctuations of the asymptotic boundaries. Indeed, when one carefully formulates JT theory in a de Sitter background one needs to include contributions from fluctuating spacetime boundaries; the resulting Schwarzian theory has attracted considerable interest recently in the context of the AdS/CFT correspondence with a microscopic realisation via the Sachdev-Ye-Kitaev model. The asymptotic symmetries of the Schwarzian theory are realised by the group $(\diff_+(S^1)\times \diff_+(S^1)/\PSL(2,\mathbb{R})$. \cite{Cotler2020_LowdimdS} 

In this section we propose a tessellation realisation of asymptotically de Sitter spacetime, and highlight the role of Thompson's group $T$ as a candidate for an analogue of the corresponding asympototic symmetries. We extend the action of $\PSL(2,\mathbb{Z})$ on our tessellation to the action of Thompson's group $T$. As a consequence, we then \emph{define} a nearly-$\text{dS}_2$ tessellation to be the resulting holographic network.

\subsection{Thompson's group $T$: tessellations for nearly $\text{dS}_2$}
Thompson's group $T$ is a remarkable group of homeomorphisms of the circle. The role of Thompson's group $T$ as a discretised analogue of $\diff(S^1)$ compatible with tessellations of $S^1$ was emphasised by Vaughan Jones \cite{jones_unitary_2014}, with later work supporting this idea in the context of the AdS/CFT correspondence \cite{Osborne2017_DynHolCode,osborne_quantum_2019}. Following these proposals, we exploit Thompson's group $T$ as an analogue of $\diff(S^1)$ in the context of tessellations of dS. As we argue, this allows for a microscropic realisation of a $\text{dS}_{1+1}/\text{TN}_{1}$ correspondence.

Here we recall the definition of Thompson's group $T$; we follow \cite{cannon_introductory_1996} and \cite{belk_thompsons_2007}. 

\begin{definition}
	\textit{Thompson's group $T$} is given by a set of piecewise linear homeomorphisms of the unit circle, with the group operation given by function composition. The elements of Thompson's group $T$ are required to satisfy the following conditions:
	\begin{itemize}
		\item they are differentiable except at finitely many points;
		\item points where they are not differentiable lie at dyadic rational numbers;
		\item when they are differentiable, the derivatives are powers of two.
	\end{itemize}
\end{definition}
It may be shown that Thompson's group $T$ is a finitely presented infinite group generated by the following three functions $A(x)$, $B(x)$, and $C(x)$ (and their inverses):
\begin{center}
	\includegraphics[scale=0.95,page=22]{figures.pdf}
\end{center}

\subsubsection{The subgroup $\PSL(2,\mathbb{Z})$ and its action on tessellations}
A remarkable fact \cite{fossas_psl2z_2010}, supporting the proposal that Thompson's group $T$ supplies an analogue of $\diff(S^1)$ appropriate for tessellations of dS, is the observation that the elements $S(x) = (C^{-1}\circ A^{-1})(x)$ and $C(x)$ generate a subgroup isomorphic to $\PSL(2,\mathbb{Z})$. These two elements are defined via 
\begin{equation}
	S(x) = \begin{cases}
		x + \frac12 , \, x < \frac12\\
		x - \frac12 , \, \frac12 \leq x 
	\end{cases}
\end{equation}
and
\begin{equation}	
	C(x) = \begin{cases}
		\frac12 \, x + \frac34 , \, x \leq \frac12\\
		2x-1, \, \frac12 \leq x \leq \frac34\\
		x-\frac14, \, \frac34 \leq x 
	\end{cases}
\end{equation}
with function graphs 
\begin{center}
	\includegraphics[page=23]{figures.pdf}
\end{center}
As one may readily observe, these two elements are precisely given by
\begin{align}
	S(x) &= \psi\circ ?\circ -\frac{1}{x}\circ ?^{-1}\circ \psi^{-1} \nonumber \\ 
	\text{and}\quad C(x) &= \psi\circ ?\circ \frac{x-1}{x}\circ ?^{-1}\circ \psi^{-1}.
\end{align}
It may be graphically verified that the functions $S(x)$ and $C(x)$ also fulfil the relations presented in Eq.~(\ref{eq:repPSL}). To describe this we exploit a representation of the functions $S(x)$ and $C(x)$ via \emph{rectangle diagrams} \cite{cannon_introductory_1996}. Here the top of the rectancle is identified with the domain of the function and the bottom with the range. Lines are drawn from top to bottom, indicating the image of given points in the domain. The action is then inferred by linearly interpolating the action between the lines.  By stacking multiple rectangle diagrams composition of functions in Thompson's group $T$ may be computed. The elements $S$ and $C$ are represented by the following rectangle diagrams:
\begin{center}
	\includegraphics[page=24]{figures.pdf}
\end{center}
Exploiting this graphical representation the identities 
\begin{align}
	(S\circ S)(x) = x \quad \text{and} \quad (C\circ C\circ C)(x) =x \label{eq:compSC}
\end{align}
may be verified:
\begin{center}
	\begin{tikzpicture}[scale = 0.5]
		\node[left,text=white] at (0,1) {$(C \circ C \circ C)(x) =$};
		\node[left] at (0,1) {$(S\circ S)(x) = $};
		\foreach \y in {0,1} {
			\draw[lightgray] (5,\y+1) -- (5,\y);
			\draw (0, \y) rectangle (10,\y);
			\draw (10,\y+1) -- (5,\y);
			\draw (5,\y+1) -- (0,\y);
		}
		\draw (0, 0) rectangle (10,2);
	\end{tikzpicture}\vspace{5mm}
	\begin{tikzpicture}[scale = 0.5]
		\node[left,text=white] at (0,1) {$(C \circ C \circ C)(x) =$};
		\node[left] at (0,1) {$ = $};
		\foreach \y in {0,1} {
			\draw[] (5,\y+1) -- (5,\y);
		}
		\draw (0, 0) rectangle (10,2);
	\end{tikzpicture}\\\vspace{8mm}
	
	\begin{tikzpicture}[scale = 0.5]
		\node[left] at (0,1.5) {$(C\circ C\circ C)(x) = $};
		\foreach \y in {0,1,2} {
			\draw[lightgray] (5,\y+1) -- (5,\y);
			\draw[lightgray] (7.5,\y+1) -- (7.5,\y);
			\draw[] (0, \y) rectangle (10,\y+1);
			\draw (5,\y+1) -- (0,\y);
			\draw (7.5,\y+1) -- (5,\y);
			\draw (10,\y+1) -- (7.5,\y);}
	\end{tikzpicture}\vspace{5mm}
	\begin{tikzpicture}[scale = 0.5]
		\node[left,text=white] at (0,1) {$(C \circ C \circ C)(x) =$};
		\node[left] at (0,1.5) {$ = $};
		\draw (0, 0) rectangle (10,3);
		\draw (5,3) -- (05,0);
		\draw (7.5,3) -- (7.5,0);
	\end{tikzpicture}
\end{center}

\begin{figure}
	\begin{subfigure}{.48\textwidth}\begin{center}
			\includegraphics[page=5]{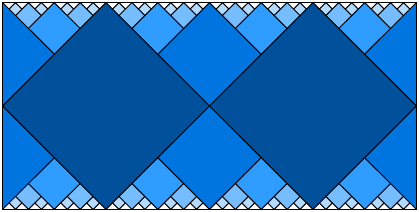}
			\caption{Tessellation after action of the generator $S(x)$}
			\label{fig:TessTransS}
	\end{center}\end{subfigure}
	\begin{subfigure}{.48\textwidth}\begin{center}
			\includegraphics[page=26]{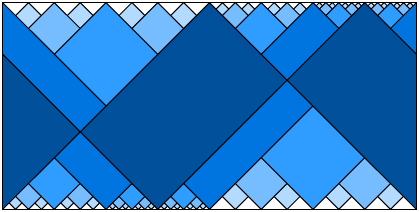}			
			\caption{Tessellation after action of the generator $C(x)$}
			\label{fig:TessTransC}
	\end{center}\end{subfigure}
	\caption{Tessellations after the action of transformations from the subgroup $\PSL(2,\mathbb{Z})$.}
	\label{fig:Tesstrans}
\end{figure}

Now we have identified the role played by $S$ and $C$ we can define the action of $\PSL(2,\mathbb{Z})$ on our tessellations. We do this by transforming the points $x_+ \in \mathcal{I^+}$ and $x_- \in \mathcal{I^-}$ defining a causal diamond according to (note that $\theta_{-\pi}=\theta_{\pi} = S$)
\begin{subequations}\begin{align}
		x_+ \rightarrow S(x_+),\quad
		x_- &\rightarrow (S \circ S \circ S) (x_-) = S(x_-),\\
		x_+ \rightarrow C(x_+),\quad
		x_- &\rightarrow (S \circ C \circ S) (x_-).
\end{align}\end{subequations}
Applying this action to all the causal diamonds defining our tessellation yields, in the case of $S$ and $C$, the transformed tessellations shown in Fig.~\ref{fig:Tesstrans}. The full action of $\PSL(2,\mathbb{Z})$ is obtained analogously. Although the generator $S(x)$ seemingly has no effect on the initial tessellation, this is not the case: it acts as an involution by mapping each point to the opposite side of the spacetime. 

The action of the generator $C(x)$ on our tessellation is depicted in Fig.~\ref{fig:TessTransC}. (Even though only a finite number of tiles are depicted, one must imagine that the tessellation actually extends to infinity.) As one may observe, this transformation yields a tessellation with tiles distorted along null geodesics. The distortion arises because the induced transformations on the temporal boundaries $\mathcal{I^+}$ and $\mathcal{I^-}$ are not identical. In this way causal diamonds are both translated and distorted. Note, however, that only a \emph{finite} number tiles are distorted by the transformation. The remaining tiles are left invariant, and the tessellation is hence \emph{almost} invariant. This residual action has consequences for the construction of the physical Hilbert space, and supplies additional constraints on the tensors comprising the holographic network to ensure that the resulting quantum mechanic system is invariant under the full isometry group $\PSL(2,\mathbb{Z})$.

A useful mnemonic to understand the action of $\PSL(2,\mathbb{Z})$ is to think of the tessellation itself as ``representing'' the process which propagates information, carried by null geodesics, around $\text{dS}_2$ from the past to the future boundaries. Thinking of this process as a function $f:\mathcal{I}^{-}\rightarrow \mathcal{I}^+$ we should, according to an isometry, e.g., $C\in \PSL(2,\mathbb{Z})$, transform the function $f$ according to
\begin{equation}
	f\mapsto C\circ f \circ (S\circ C^{-1}\circ S).
\end{equation}
In the case $f = S$ the action is trivial on $f$. However, the action is nontrivial in general on tessellations and, as we'll see, on the physical Hilbert space $\mathcal{H}_{\text{phys}}$. 

\subsubsection{The action of Thompson's group $T$: tessellations of nearly $\text{dS}_2$}
Here we describe how to extend the action of $\PSL(2,\mathbb{Z})$ described above to an action of Thompson's group $T$. We propose that the resulting tessellations should be thought of as discretised versions of nearly $\text{dS}_2$.

The definition of an asymptotic symmetry of a spacetime, such as $\text{dS}_2$, is nontrivial. Here we broadly follow the book of Wald \cite{Wald1984} and the thesis of Jäger \cite{jager2008conserved}, we use the definition provided in \cite{jager2008conserved}. However, since we are only using these notions as a \emph{motivation} for our definition, we won't dwell on the subtle details. It suffices, as motivation, to say that asymptotic symmetries of nearly $\text{dS}_2$ are determined by diffeomorphisms of the boundaries $\mathcal{I}^{\pm}$, which are conformal with respect to the induced boundary metric. In our case, the conformality condition is trivial (our boundary manifolds are one dimensional), and we understand asymptotic symmetries as diffeomorphisms of the past and future boundaries. We should mod out by all isometries to find those symmetries which are truly nontrivial, hence one finds that the group of asymptotic symmetries of $\text{dS}_2$ is given by $(
\diff(S^1)\times \diff(S^1))/\PSL(2,\mathbb{R})$.

Employing the idea that $T$ is the correct substitute for $\diff(S^1)$ in the context of a tessellation leads us to postulate that the correct group to describe asymptotic symmetries of the tessellation is hence
\begin{equation}
	G \equiv (T\times T)/\PSL(2,\mathbb{Z}).
\end{equation}
This group acts as follows. Given a representative $(f,g) \star (h,S\circ h\circ S)$, where $f,g\in T$ and $h\in \PSL(2,\mathbb{Z})$, and the group operation $\star$ is elementwise composition
\begin{equation}
	(f,S\circ g\circ S)\star(h,S\circ h\circ S) \equiv (f\circ h, S\circ g\circ h\circ S),
\end{equation}
we obtain a new tessellation by applying $f\circ h$ to the future boundary and $S\circ g\circ h\circ S$ to the past boundary. This operation shifts the endpoints of causal diamonds, leaving us with a new tessellation. 
Commencing with the original tessellation, one can generate via this action a family of tessellations, and thereby, tensor networks, corresponding to elements of the asymptotic symmetry group. 

\subsection{Transforming holographic networks}

Given a symmetry transformed tessellation one can directly construct the corresponding holographic network according to Sec.~\ref{sec:TNconstruct1}: tensors are placed at the corners and legs from neighbouring generations are contracted to build a tensor network. We write $M^f$ for the tensor network arising in this way from the action of $f\in \PSL(2,\mathbb{Z})$. 

To explain the construction of $M^f$ we focus on the element $C\in \PSL(2,\mathbb{Z})$, with the generalisation to arbitrary $f\in \PSL(2,\mathbb{Z})$ left to the reader. Recall from Sec.~\ref{subsubsec:kinhilbertspace} that $\mathcal{H}_{\text{in}}$ should be thought of as the ``limit'' $n\rightarrow \infty$ of the finite-dimensional \emph{cutoff} Hilbert spaces $\mathcal{H}_{-n} \equiv \bigotimes_{j=0}^{2^n-1}\mathbb{C}^D$. In this way the operator $M$ is really thought of as a family of linear operators $M_{m,n}$ mapping from the cutoff initial Hilbert space $\mathcal{H}_m$ associated to a time slice $T_m$ to the cutoff final Hilbert space $\mathcal{H}_n$ associated with the time slice $T_n$. An explicit example is the tensor network for $\ell = 2$:
\begin{center}
	\includegraphics[page=13]{figures.pdf}
\end{center}
Here $m = -4$ and $n= 4$, i.e., $M_{-4,4}: (\mathbb{C}^{D})^{\otimes 32} \rightarrow (\mathbb{C}^{D})^{\otimes 32}$. Acting with $C\in \PSL(2,\mathbb{Z})$ and building the resulting tensor network yields the following operator:
\begin{center}
	\includegraphics[page=27]{figures.pdf}
\end{center}
This operator still has 32 input and output legs. \emph{However}, due to the distortion induced by $C$ on the past and future boundaries, these legs are no longer associated to a regular cutoff. That is, the transformation produces a nonregular grid. To proceed we exploit the fact that this operator is really just a member of an infinite family of physically equivalent operators acting on arbitrary cutoffs. We obtain an equivalent operator compatible with a regular (but finer) cutoff via a ``partial UV completion'' where we add in the tensors (shown in blue) necessary to define an operator which does act on a regular grid:
\begin{center}
	\includegraphics[page=28]{figures.pdf}
\end{center}
In this case we have described this network as an operator $M^C_{m',n'}$ with $m' = -6$ and $n'=6$. (It is possible to give a more compact description with $m''=-4$ and $n''=4$ by removing a layer of isometries; this would then be the most compact description.) Note that the action of the isometry entails an effective change of scale; it is often not possible to define a transformed tensor network $M^f$ while retaining the original cutoff. The final step to defining $M^C$ is then to take the limit $m'\rightarrow -\infty$ and $n'\rightarrow \infty$, i.e.,
\begin{equation}
	M^C \text{``$=$''} \lim_{\substack{m'\rightarrow -\infty\\ n'\rightarrow \infty}} M^C_{m',n'}.
\end{equation}

It turns out that one can express the transformed operator $M^f$ as
\begin{equation}
	M^f \equiv U(f)M U^{\dag}(\theta_{-\pi}\circ f \circ \theta_\pi),
\end{equation}
where $U(f)$ is a unitary operator whose action we now describe. (The following discussion is an abbreviated version of the discussion in Sec.~4.4 of \cite{Osborne2017_DynHolCode}, and the reader is invited to skim over that paper for a more comprehensive discussion of unitary representations of Thompson's group $T$.) We focus again on the case $f = C$ for simplicity and describe how to construct the unitary operator $U(C):\mathcal{H}_{\text{in}}\rightarrow \mathcal{H}_{\text{in}}$. Intuitively $U(C)$ is just the operator which moves the qudits comprising $\mathcal{H}_{\text{in}}$ around according to the function $C$. E.g., a qudit at location $\theta = \frac{3}{4}\times 2\pi$ gets moved to $\theta = \frac{1}{2}\times 2\pi$, etc. To actually describe this action we choose an element $|\phi_m\rangle$ of $\mathcal{H}_{m}\subset\mathcal{H}_{\text{in}}$. (Recall that, as described in Sec.~\ref{subsubsec:kinhilbertspace}, two elements $|\psi_m\rangle$ and $|\psi'_{m'}\rangle$ with $m'<m$ are equivalent in $\mathcal{H}_{\text{in}}$ if $M_{m',m}|\psi'_{m'}\rangle = |\psi_m\rangle$.) We, for concreteness, set $m = -2$ so that $|\phi\rangle \in (\mathbb{C}^{D})^{\otimes 4}$. This initial state is thought of as that of a quantum spin system with the $4$ spins regularly distributed around the circle: 
\begin{center}
	\begin{tikzpicture}[scale=0.5]
		\draw (0,0) -- (16,0);
		\foreach \x in {0,4,...,16}
		{
			\draw (\x,-0.2) -- (\x, 0.2);
		}
	\end{tikzpicture}
\end{center}
After applying $C$ the $4$ spins are moved around and dilated/contracted: e.g., the second spin is now associated to the last interval with half its original length, and the third spin is now the first spin of an interval of twice its original length. Thus we must now assign this permuted state $|\phi'\rangle$ of the spins to a \emph{nonregular grid}:
\begin{center}
	\begin{tikzpicture}[scale=0.5]
		\draw (0,0) -- (16,0);
		\foreach \x in {0,8}
		{
			\draw (\x,-0.2) -- (\x, 0.2);
		}
		\foreach \x in {12}
		{
			\draw (\x,-0.2) -- (\x, 0.2);
		}
		\foreach \x in {14,16}
		{
			\draw (\x,-0.2) -- (\x, 0.2);
		}
	\end{tikzpicture}
\end{center}
This nonregular grid is no longer comparable with the original regular grid: to find a state of a regular grid we need to fine-grain the state $|\phi'\rangle$ of the nonregular grid using the $V$ tensor:
\begin{equation}
	|\phi''\rangle \equiv [(V\otimes V)V] \otimes V\otimes \mathbb{I}|\phi'\rangle
\end{equation}
This new state is now associated to a regular grid with $8$ intervals:
\begin{center}
	\begin{tikzpicture}[scale=0.5]
		\foreach \x in {2,4,...,6}
		{
			\draw[color3] (\x,-0.2) -- (\x, 0.2);
		}
		\draw[color3] (10,-0.2) -- (10, 0.2);
		\draw (0,0) -- (16,0);
		\foreach \x in {0,8}
		{
			\draw (\x,-0.2) -- (\x, 0.2);
		}
		\foreach \x in {12}
		{
			\draw (\x,-0.2) -- (\x, 0.2);
		}
		\foreach \x in {14,16}
		{
			\draw (\x,-0.2) -- (\x, 0.2);
		}

	\end{tikzpicture}
\end{center}
Thus $U(C)$ acts on elements of $\mathcal{H}_{-2}$ via
\begin{equation}
	|\phi\rangle \overset{U(C)}{\mapsto} [(V\otimes V)V] \otimes V\otimes \mathbb{I}|\phi'\rangle,
\end{equation}
where $|\phi'\rangle$ is the cyclically permuted version of $|\phi\rangle$. This action is, in general, nontrivial. To see this we compare the original state to the transformed version $|\phi''\rangle \equiv U(C)|\phi\rangle$, i.e., we would like to compute the inner product
\begin{equation}
	\langle \phi | U(C)|\phi\rangle
\end{equation}
This is not possible within the Hilbert space of the original grid of $4$ spins. Instead we must first fine-grain the original state $|\phi\rangle$ to an equivalent state of a grid of $8$ spins and then compute the overlap:
\begin{equation}
	\begin{split}
		\langle \phi | U(C)& |\phi\rangle \\ \equiv& \langle\phi|(V^{\dag}\otimes V^{\dag}\otimes V^{\dag}\otimes V^{\dag})[(V\otimes V)V] \otimes V\otimes \mathbb{I}|\phi'\rangle \\
	=& \langle\phi|V\otimes \mathbb{I}\otimes V^{\dag}|\phi'\rangle.
	\end{split}
\end{equation}
Depending on $|\phi\rangle$ and $V$ this overlap may be nontrivial and the action is nontrivial. The action of $U(f)$ for $f\in T$ on any state in $\mathcal{H}_{-m}$ may be computed in a similar fashion and furnishes a unitary representation of Thompson's group $T$ on $\mathcal{H}_{\text{in}}$. (An analogous construction works on $\mathcal{H}_{\text{out}}$.)

It now may be argued that, with this unitary representation in hand, the tensor network corresponding to a tessellation transformed by $f\in T$ is given by
\begin{equation}
	M^f \equiv U(f)M U^{\dag}(S)U^\dag(f)U(S)^\dag.
\end{equation}
(Here we have used the fact that $\theta_{\pi} = S$ and the fact that $U(f) U(g) = U(f\circ g)$.) This result allows us to rapidly deduce that the transformed tensor network $M_f$ is a partial isometry from $\mathcal{H}_{\text{in}}$ to $\mathcal{H}_{\text{out}}$. 

Note that, when the building-block tensors $U$ and $V$ in Eq.~(\ref{eq:uandvtensors}) are arbitrary, there is no reason to expect that $M=M^f$. This means that, in general, the family of projections
\begin{equation}
	P_f \equiv (M^f)^\dag M^f
\end{equation}
acting on $\mathcal{H}_{\text{in}}$ project onto differing subspaces $\mathcal{H}_{\text{phys}}^f \equiv P_f \mathcal{H}_{\text{in}}$. Either one regards this a gauge degree of freedom, i.e., one posits that the subspaces $\mathcal{H}_{\text{phys}}^f$ are gauge equivalent, or one demands that 
\begin{equation}\label{eq:qeccgaugecond}
	\mathcal{H}_{\text{phys}}^f = \mathcal{H}_{\text{phys}}
\end{equation}
for all $f\in\PSL(2,\mathbb{Z})$. This latter point of view imposes additional constraints on $U$ and $V$.

The conditions on $U$ and $V$ so that Eq.~(\ref{eq:qeccgaugecond}) exactly holds are not presently clear, and we haven't yet assembled enough information to formulate a precise conjecture. Instead we find a class of examples where one can argue that 
\begin{equation}
	\mathcal{H}_{\text{phys}}' \equiv \bigcap_{f\in\PSL(2,\mathbb{Z})} \mathcal{H}_{\text{phys}}^f \not= \emptyset.
\end{equation}
These examples are furnished in any braided fusion category by choosing $U$ to be the braiding operation
\begin{center}
	\includegraphics[page=34]{figures.pdf}
\end{center}
and the isometry $V$ to satisfy the pivotality condition
\begin{center}
	\includegraphics[page=35]{figures.pdf}
\end{center}
For such tensors one can construct an invariant state $|\Omega\rangle \in \mathcal{H}_{\text{in}}$ as the infinite regular binary tree
\begin{center}
	\includegraphics[page=25, width=\columnwidth]{figures.pdf}
\end{center}
Invariance of $|\Omega\rangle$ under $M$ follows directly from the isometry conditions for $U$ and $V$ and pivotality:
\begin{center}
	\includegraphics[page=31, width=\columnwidth]{figures.pdf}
\end{center}
Invariance under the action of $C$ relies on the braiding and pivotality conditions:
\begin{center}
	\includegraphics[page=32, width=\columnwidth]{figures.pdf}
\end{center}
which simplifies to $|\Omega\rangle$ thanks to the following sequence of moves
\begin{center}
	\includegraphics[page=33, width=\columnwidth]{figures.pdf}
\end{center}
Thus we have shown that $\mathcal{H}'_{\text{phys}}$ is nonempty. 

\section{Conclusions and outlook} 
In this paper we have proposed a family of holographic tensor networks associated to tessellations of de Sitter  spacetime in $(1+1)$ dimensions. These tensor networks take the form of \emph{partial isometries} mapping from an infinite-dimensional kinematical Hilbert space $\mathcal{H}_{\text{in}}$ associated to the negative temporal boundary $\mathcal{I}^{-}$ to a kinematical Hilbert space $\mathcal{H}_{\text{out}}$ associated to $\mathcal{I}^{+}$. The associated projection on $\mathcal{H}_{\text{in}}$ has finite rank and singles out a finite-dimensional subspace of \emph{physical states} of $\mathcal{H}_{\text{in}}$. The de Sitter radius is modelled by subdividing the initial tessellation and implies a direct connection between the quantum information carrying capacity of the tensor network and the cosmological constant. The action of isometries on the network and resulting Hilbert space was described, along with conditions on the constituent tensors ensuring that (a subspace of) the physical Hilbert space is invariant. Finally, asymptotic symmetries were identified with Thompson's group $T$. 

This paper only just barely scratches the surface of possibilities. Many fascinating and challenging problems remain. A partial list of intriguing questions includes
\begin{itemize}
	\item Quantum error correction and dS: the holographic tensor networks we've described here are related to \emph{concatenated quantum error correcting codes}. This resemblence is only superficial at the present stage, and to make more direct contact one requires tessellations leading to higher-order tensors (e.g., involving $6$-leg tensors). Such a generalization is natural and should enable results concerning bulk reconstruction. 
	\item Bulk reconstruction: can one reconstruct bulk perturbations at the temporal boundaries? This would probably require Witten's \emph{superobservers}, but perhaps employing ideas from quantum error correction could help here.
	\item The tensor networks proposed here enjoy certain degenerate entropy/area relationships. Can one microscopically elucidate a general Ryu-Takayanagi formula for dS?
	\item Higher dimensions: We have only considered $(1+1)$ dimensions here. It is possible to define analogous tensor networks in higher dimensions for noncompact spatial manifolds. Regular tessellations for compactified spatial manifolds are impossible, and new ideas are needed here.
	\item Black holes: can one associate tensor networks to Schwarzschild de Sitter solutions? 
	\item Wheeler de Witt: there is a candidate tensor network for the Wheeler de Witt solution (apply the tensor network for dS to a standard holographic code). Does this microscopic model supply us with new insights?
	\item JT gravity: is there a natural microscopic realisation of JT gravity using the nearly dS tessellations we've defined here?
	\item Quantum information capacity and the cosmological constant: is there a deeper connection between the cosmological constant and the quantum information capacity of spacetime?
\end{itemize}
We are optimistic that the microscopic perspective introduced in this paper will enable progress on some of these challenging problems.

\begin{acknowledgments}
Helpful discussions with Jordan Cotler and Deniz Stiegemann are gratefully acknowledged. This work was supported, in part, by the Quantum Valley Lower Saxony (QVLS), the DFG through SFB 1227 (DQ-mat), the RTG 1991, and funded by the Deutsche Forschungsgemeinschaft (DFG, German Research Foundation) under Germany's Excellence Strategy EXC-2123 QuantumFrontiers 390837967.
\end{acknowledgments}

\begin{appendix} 
\section{Properties of geodesics}

\subsection{Metric in global coordinates}
\label{app:metricGlobal}
We write global coordinates and obtain the induced metric of dS with respect to this coordinate system as follows
\begin{align}
	\text{Global coordinates:}\,\left(\begin{array}{c}
		x_0 \\ x_1 \\ x_2
	\end{array}\right)
	&= \left(\begin{array}{c}
		\sinh\tau \\ \cos\theta \cosh\tau \\ \sin\theta \cosh\tau
	\end{array}\right)\\ 
	\text{Metric de Sitter:} \,\hspace{7mm} \mathrm{d}s^2 &= -\mathrm{d}x_0^2 + \mathrm{d}x_1^2 + \mathrm{d}x_2^2
\end{align}
Considering the infinitesimals
\begin{align*}
	\mathrm{d}x_0 &= \frac{\mathrm{d}x_0}{\mathrm{d}\tau}\mathrm{d}\tau + \frac{\mathrm{d}x_0}{\mathrm{d}\theta}\mathrm{d}\theta = \cosh\tau \,\mathrm{d}\tau\\
	\mathrm{d}x_1 &= \frac{\mathrm{d}x_1}{\mathrm{d}\tau}\mathrm{d}\tau + \frac{\mathrm{d}x_1}{\mathrm{d}\theta}\mathrm{d}\theta = \cos\theta \sinh\tau \,\mathrm{d}\tau - \sin\theta \cosh\tau\, \mathrm{d}\theta\\
	\mathrm{d}x_2 &= \frac{\mathrm{d}x_2}{\mathrm{d}\tau}\mathrm{d}\tau + \frac{\mathrm{d}x_2}{\mathrm{d}\theta}\mathrm{d}\theta = \sin\theta \sinh\tau \,\mathrm{d}\tau + \cos\theta \cosh\tau\, \mathrm{d}\theta 
\end{align*}
we obtain
\begin{align}
	\mathrm{d}s^2 =& 
	-\cosh^2\tau \,\mathrm{d}\tau^2 + \left(\cos\theta \sinh\tau \,\mathrm{d}\tau - \sin\theta \cosh\tau\, \mathrm{d}\theta\right)^2 \nonumber\\&
	+ \left(\sin\theta \sinh\tau \,\mathrm{d}\tau + \cos\theta \cosh\tau\, \mathrm{d}\theta \right)^2\nonumber\\
	=& 
	-\cosh^2\tau \,\mathrm{d}\tau^2 + \cos^2\theta \sinh^2\tau \,\mathrm{d}\tau^2 + \sin^2\theta \cosh^2\tau\, \mathrm{d}\theta^2\nonumber\\& + \sin^2\theta \sinh^2\tau \,\mathrm{d}\tau^2 
	+ \cos^2\theta \cosh^2\tau\, \mathrm{d}\theta^2\nonumber\\
	=& 
	\left( -\cosh^2\tau + \cos^2\theta \sinh^2\tau + \sin^2\theta \sinh^2\tau\right)\mathrm{d}\tau^2 \nonumber\\&
	+ \left( \sin^2\theta \cosh^2\tau + \cos^2\theta \cosh^2\tau\right)\mathrm{d}\theta^2\nonumber\\
	=& -\mathrm{d}\tau^2 + \cosh^2\tau \mathrm{d}\theta^2\nonumber\\
	=& g_{\mu\nu} \mathrm{d}x^\mu \mathrm{d}x^\nu
\end{align}
In matrix form:
\begin{align}
	g_{\mu\nu} =  \left(\begin{array}{c c}
		-1 & 0 \\ 0 & \cosh^2\tau
	\end{array}\right)\quad \text{and}\quad
	g^{\mu\nu} =  \left(\begin{array}{c c}
		-1 & 0 \\ 0 & \frac1{\cosh^2\tau}
	\end{array}\right).
\end{align}

\subsection{Geodesics in de Sitter spacetime embedded in Minkowski spacetime} \label{app:geodGlobal} 

The geodesics of de Sitter spacetime embedded in Minkowski spacetime can be parametrised using global coordinates:
\begin{align}
	\begin{pmatrix}
		x_0 \\x_1 \\ x_2	
	\end{pmatrix}	 = \begin{pmatrix}
		\sinh(\tau) \\ \sin(\theta) \pm \cos(\theta) \sinh(\tau) \\ \cos(\theta) \mp \sin(\theta) \sinh(\tau)
	\end{pmatrix} \label{eq:geodesicGlobal}
\end{align}

We consider the case of the anticlockwise geodesic. Geodesics embedded in Minkowski spacetime are defined in (\ref{eq:geodesic}). The anticlockwise geodesicis is
\begin{equation}
	x_\text{ac}(s)
	=\begin{pmatrix}s \\ u + vs \\ v - us\end{pmatrix} \quad \Rightarrow \quad 
	\begin{array}{r l}
		x_{\text{ac},0} &= s\\
		x_{\text{ac},1} &= u + v s\\
		x_{\text{ac},2} &= v - us
	\end{array}
\end{equation}
The parameters $u$ and $v$
($u^2+v^2=1$)
have the following properties such that the geodesic fulfils the hyperboloid condition from Eq.~(\ref{eq:hyperboloidMatr}):
\begin{equation}
	u = \frac{x_{\text{ac},1} - x_{\text{ac},0} x_{\text{ac},2}}{1+x_{\text{ac},0}^2} \quad \text{}\quad 
	v = \frac{x_{\text{ac},2} + x_{\text{ac},0} x_{\text{ac},1}}{1+x_{\text{ac},0}^2} \quad \text{}
\end{equation}
We want to show that the proposed anticlockwise geodesic from Eq.~(\ref{eq:geodesicGlobal}) actually is a geodesic
\begin{align}
	x_{\text{ac}} = \begin{pmatrix}
		\sinh(\tau) \\ \sin(\theta) + \cos(\theta) \sinh(\tau) \\ \cos(\theta) - \sin(\theta) \sinh(\tau)
	\end{pmatrix} = \begin{pmatrix}
		x_0\\x_1\\x_2
	\end{pmatrix}
\end{align}
This is solved for $\cos(\theta)$ and $\sin(\theta)$. They are identified with the parameters $u$ and $v$. 
We know that $x_0=\sinh\tau$.
\begin{align*}
	x_1=\sin\theta + \cos\theta \sinh\tau \quad \Leftrightarrow\quad \sin\theta = x_1 - \cos\theta\sinh\tau\\
	x_2=\cos\theta - \sin\theta \sinh\tau \quad \Leftrightarrow\quad \cos\theta = x_2 + \sin\theta \sinh\tau
\end{align*}
Thus
\begin{minipage}{0.5\textwidth}
	\begin{align*}
		x_1 &= \sin\theta + \left(x_2+\sin\theta\sinh\tau\right)\sinh\tau\\
		&= \sin\theta + x_2 x_0 + \sin\theta x_0^2
	\end{align*}
	\begin{align*}
		\Rightarrow \sin\theta = \frac{x_1-x_2 x_0}{1+x_0^2}
	\end{align*}
\end{minipage}
\begin{minipage}{0.5\textwidth}
	\begin{align*}
		x_2&=\cos\theta - \left(x_1 - \cos\theta\sinh\tau\right) \sinh\tau\\
		&= \cos\theta - x_0 x_1 + \cos \theta x_0^2
	\end{align*}
	\begin{align*}
		\Rightarrow \cos\theta = \frac{x_2+x_0x_1}{1+x_0^2}
	\end{align*}
\end{minipage}\\

With $u=\sin\theta$ and $v=\cos\theta$ this fulfils the condition $u^2+v^2=1$.

\subsection{Geodesics in global and conformal coordinates} 
\label{app:geodesicProof}
In this appendix we explicitly verify that the null geodesics described in this paper satisfy the geodesic equation. These calculations are elementary, however we repeat them here for completeness.

First, the geodesic equation is derived in global coordinates. Therefore, the Christoffel symbols need to be calculated:

\begin{align}
	\Gamma _{{\mu }{\nu }}^{\sigma }=\frac {1}{2}g^{{\sigma }{\kappa }}\left({\frac {\partial g_{{\nu }{\kappa }}}{\partial x^{\mu }}}+{\frac {\partial g_{{\mu }{\kappa }}}{\partial x^{\nu }}}-{\frac {\partial g_{{\mu }{\nu }}}{\partial x^{\kappa }}}\right)
\end{align}
When calculating the Christoffel Symbols, it can be used immediately, that the off diagonal elements of the metric are zero. The only derivative that does not vanish is $\frac{\partial g_{\theta\theta}}{\partial \tau} = \sinh(\tau)$. We get the following Christoffel Symbols:
\begin{align*}
	\Gamma^\tau{}_{\tau\tau} &= \frac {1}{2}g^{\tau \tau}\left({\frac {\partial g_{\tau \tau}}{\partial \tau}}+{\frac {\partial g_{\tau\tau}}{\partial\tau}}-{\frac {\partial g_{\tau\tau}}{\partial \tau}}\right)=0\\
	\Gamma^\tau{}_{\theta\tau} &= \Gamma^\tau{}_{\tau\theta} = \frac {1}{2}g^{\tau \tau}\left({\frac {\partial g_{\tau\tau}}{\partial\theta}}+{\frac {\partial g_{\theta \tau}}{\partial \tau}}-{\frac {\partial g_{\theta\tau}}{\partial \tau}}\right)=0\\
	\Gamma^\tau{}_{\theta\theta} &= \frac {1}{2}g^{\tau \tau}\left({\frac {\partial g_{\theta \tau}}{\partial \theta}}+{\frac {\partial g_{\theta\tau}}{\partial\theta}}-{\frac {\partial g_{\theta\theta}}{\partial \tau}}\right) \\&
	=\frac{1}{2} (-1) \left(-\frac{\partial \cosh^2\tau}{\partial\tau}\right) 
	= \cosh\tau \sinh\tau\\
	\Gamma^\theta{}_{\theta\theta} &= \frac {1}{2}g^{\theta \theta}\left({\frac {\partial g_{\theta \theta}}{\partial \theta}}+{\frac {\partial g_{\theta\theta}}{\partial\theta}}-{\frac {\partial g_{\theta\theta}}{\partial \theta}}\right)=0\\
	\Gamma^\theta{}_{\theta\tau} &= \Gamma^\theta{}_{\tau\theta} = \frac {1}{2}g^{\theta \theta}\left({\frac {\partial g_{\tau\theta}}{\partial\theta}}+{\frac {\partial g_{\theta \theta}}{\partial \tau}}-{\frac {\partial g_{\theta\tau}}{\partial \theta}}\right) \\&
	=\frac{1}{2}\frac{1}{\cosh^2\tau} \left(\frac{\partial \cosh^2\tau}{\partial \tau}\right)
	=  \frac{ \sinh\tau}{\cosh\tau} = \tanh\tau \\
	\Gamma^\theta{}_{\tau\tau} &= \frac {1}{2}g^{\theta \theta}\left({\frac {\partial g_{\tau \theta}}{\partial \tau}}+{\frac {\partial g_{\tau\theta}}{\partial\tau}}-{\frac {\partial g_{\tau\tau}}{\partial \theta}}\right)=0\\
\end{align*}

The general form of the geodesic equation is
\begin{align}
	\frac{\mathrm{d}^2x^\lambda}{\mathrm{d}t^2} + \Gamma^\lambda{}_{\mu\nu} \frac{\mathrm{d}x^\mu}{\mathrm{d}t} \frac{\mathrm{d}x^\nu}{\mathrm{d}t}=0
\end{align}

The geodesic equations for $\tau$ and $\theta$ can be written down:
\begin{align}
	\frac{\mathrm{d}^2 \tau(t)}{\mathrm{d}t^2} + \Gamma^\tau{}_{\theta\theta} \frac{\mathrm{d} \theta(t)}{\mathrm{d}t} \frac{\mathrm{d} \theta(t)}{\mathrm{d}t} &=0\nonumber\\
	\Rightarrow\quad  \frac{\mathrm{d}^2 \tau(t)}{\mathrm{d}t^2} +  \cosh\tau(t) \sinh\tau(t) \left( \frac{\mathrm{d} \theta(t)}{\mathrm{d}t} \right)^2 &=0  \label{eq:GEtau}\\
	\frac{\mathrm{d}^2 \theta(t)}{\mathrm{d}t^2} + 2 \, \Gamma^\theta{}_{\theta\tau} \frac{\mathrm{d} \theta(t)}{\mathrm{d}t} \frac{\mathrm{d} \tau(t)}{\mathrm{d}t} &=0\nonumber\\
	\Rightarrow\quad  \frac{\mathrm{d}^2 \theta(t)}{\mathrm{d}t^2} + 2 \tanh\tau(t) \frac{\mathrm{d} \theta(t)}{\mathrm{d}t} \frac{\mathrm{d} \tau(t)}{\mathrm{d}t} &=0  \label{eq:GEtheta}
\end{align}
This geodesic equation is solved by the following null geodesic in global coordinates:
\begin{align}
	\begin{pmatrix}
		\theta(t) \\ \tau(t) 
	\end{pmatrix} = \begin{pmatrix}
		\pm \arctan t + \theta_0 \\ \arcsinh t 
	\end{pmatrix} \qquad t\in \mathbb{R}
\end{align}

This null geodesic in global coordinates can be transformed to conformal coordinates:
\begin{align*}
	\cos(T) &= \frac{1}{\cosh(\tau)} \qquad \Rightarrow \qquad T = \pm \arccos\left(\frac{1}{\cosh \tau}\right)
\end{align*}
The resulting null geodesic in conformal coordinates is
\begin{align}
	\begin{pmatrix}
		\theta(t) \\ T(t) 
	\end{pmatrix} &= \begin{pmatrix}
		\pm \arctan t + \theta_0\\ \pm \arccos\left(\frac{1}{\cosh (\arcsinh t )}\right)
	\end{pmatrix} \nonumber\\&= 
	\begin{pmatrix}
		\pm \arctan t + \theta_0\\ \pm \arccos \left(\frac{1}{\sqrt{t^2+1}}\right)
	\end{pmatrix}
\end{align}

The parameter $t$ can be substituted with $t=\tan s$ with $-\frac{\pi}{2}<s<\frac{\pi}{2}$.

\begin{align}
	\begin{pmatrix}
		\theta(s) \\ T(s) 
	\end{pmatrix} &= 
	\begin{pmatrix}
		\pm \arctan (\tan s) + \theta_0\\\arccos \left(\frac{1}{\sqrt{\tan^2 s+1}}\right)
	\end{pmatrix}
\end{align}

For $-\frac{\pi}{2}<s<\frac{\pi}{2}$ this can be further simplified:
\begin{align*}
	\begin{pmatrix}
		\theta(s) \\ T(s) 
	\end{pmatrix} &= 
	\begin{pmatrix}
		\pm s + \theta_0\\
		\pm \arccos(\pm \cos s) 
	\end{pmatrix}, \qquad -\frac{\pi}{2}< s <\frac{\pi}{2}
\end{align*}
Given that
\begin{align*}
	T(s) = \pm \arccos(\pm \cos s) = \begin{cases}
		\pm\vert s \vert &,\,+\\ \pm(\pi - \vert s \vert) &,\,-
	\end{cases}
\end{align*}
we can choose one valid solution for the geodesic. Hereby we know, that the variable $T$ and the parameter $s$ lie in the boundaries $-\frac{\pi}{2}\leq T\leq \frac{\pi}{2}$. The geodesic can be described as

\begin{align}
	\begin{pmatrix}
		\theta(s) \\ T(s)
	\end{pmatrix} = \begin{pmatrix}
		\theta_0  \pm s \\ s
	\end{pmatrix}, \qquad -\frac{\pi}{2}< s <\frac{\pi}{2}
\end{align}
This yields a geodesic in conformal coordinates that is a straight line tilted at $45^\circ$.

\section{Basics on tensor networks}
\label{app:TNbasics}
Our review of the basic notions of tensor networks {follows} the paper by Bridgeman and Chubb \cite{bridgeman_hand-waving_2017}.\\

The basic building block of any tensor network is a tensor. A key defining property of a tensor is its rank: A $d$-dimensional vector is a rank-$1$ tensor which is an element of $\mathbb{C}^d$. Similarly, a $(n\times m)$-matrix is a rank-$2$ tensor which is an element of $\mathbb{C}^{n\times m}$. This motivates the definition of a rank-$r$ tensor with dimensions $d_1\times \cdots\times d_r$ which is an element of $\mathbb{C}^{d_1\times \cdots\times d_r}$. The number of indices of the tensor in index notation and the number of legs in the tensor network notation correspond to the rank of the tensor. To illustrate the tensor network notation, we look at an example of a rank-$4$ tensor:
\begin{center}
	\begin{tikzpicture}[scale=0.6]
		\node at (-1.8,0) {$=$};
		\node[left] at (-1.9,0) {$X^\rho_{\sigma\mu\nu}$};
		\draw (0,0) -- (-1.5,-1.);
		\draw (0,0) -- (1.5,-1.);
		\draw (0,0) -- (0,-1.5);
		\draw (0,0) -- (0,1.5);
		\draw[fill=white] (0,0) circle (0.5);
		\node at (0,0) {$X$};
		\node[right] at (0,1.3) {\scriptsize $\rho$};
		\node[above] at (-1.5,-0.9) {\scriptsize $\sigma$};
		\node[right] at (0,-1.4) {\scriptsize $\mu$};
		\node[above] at (1.5,-0.9) {\scriptsize $\nu$};
	\end{tikzpicture}
\end{center}

We interpret lower tensor legs as incoming legs and upper tensor legs as outgoing legs. 
The direction of the legs of the tensor is hereby associated with covariant and contravariant indices in the Einstein notation. The incoming and outgoing tensor legs are associated with different Hilbert spaces.
Each tensor is proportional to a map from the Hilbert space associated with the incoming tensor legs to the Hilbert space associated with the outgoing tensor legs.
A tensor $X^\dagger$  which is adjoint to the tensor $X$ can be expressed as follows: In the tensor network notation upper and lower legs are flipped, such that the tensor is mirrored along the constant time slice it sits on. 
\begin{center}
	\begin{tikzpicture}[scale=0.6]
		\node at (-1.8,0) {$=$};
		\node[left] at (-1.9,0) {$(X^\rho_{\sigma\mu\nu})^\dagger = X^{\sigma\mu\nu}_\rho $};
		\draw (0,0) -- (-1.5,1.);
		\draw (0,0) -- (1.5,1.);
		\draw (0,0) -- (0,1.5);
		\draw (0,0) -- (0,-1.5);
		\draw[fill=white] (0,0) circle (0.5);
		\node at (0,0) {$X^\dagger$};
		\node[right] at (0,-1.3) {\scriptsize $\rho$};
		\node[below] at (-1.5,0.9) {\scriptsize $\sigma$};
		\node[right] at (0,1.4) {\scriptsize $\mu$};
		\node[below] at (1.5,0.9) {\scriptsize $\nu$};
	\end{tikzpicture}
\end{center}

\begin{definition} An \emph{isometry} is a linear map $T:\mathcal{H}_A \rightarrow\mathcal{H}_B$ between the Hilbert spaces $\mathcal{H}_A$ and $\mathcal{H}_B$ which preserves the inner product. 
\end{definition}
A linear map $T:\mathcal{H}_A \rightarrow \mathcal{H}_B$ can be expressed as follows in index notation, where $\{\ket{a}\}$ is the complete orthonormal basis of the Hilbert space $\mathcal{H}_A$ and $\{\ket{b}\}$ is the complete orthonormal basis of the Hilbert space $\mathcal{H}_B$:
\begin{align}
	T: \ket{a} \mapsto \sum_b \ket{b} T_{ba}
\end{align}
The linear map $T$ is an isometry if and only if 
\begin{align}
	\sum_b T^\dagger_{a' b} T_{ba} = \delta_{a' a}
\end{align}
Graphically, this is represented as follows: 
\begin{center}
	\begin{tikzpicture}[scale=0.7]
		\draw (0,-2) -- (0,2);
		\draw[fill=white] (0,-1) circle (0.5);
		\draw[fill=white] (0,1) circle (0.5);
		\node at (0,-1) {$T$};
		\node at (0,1) {$T^\dagger$};
		\node at (1,0) {$=$};
		\draw (2,-2) -- (2,2);
		\node[right] at (0,0) {\scriptsize $b$};
		\node[right] at (0,-2) {\scriptsize $a$};
		\node[right] at (0,2) {\scriptsize $a'$};
		\node[right] at (2,-2) {\scriptsize $a$};
		\node[right] at (2,2) {\scriptsize $a'$};
	\end{tikzpicture}
\end{center}
\begin{definition} A \emph{perfect tensor} is a tensor $T_{a_1 \cdots a_n}$ with $n$ indices that is proportional to an isometric tensor from $A$ to $A^C$ for any bipartition of its indices into a set $A$ and its complementary set $A^C$ with $\vert A\vert \leq \vert A^C\vert$.
\end{definition}

In order to build tensor networks we introduce some basic tensor operations.
\begin{definition}The \emph{tensor product} is a generalisation of the outer product of vectors. It is defined as the element-wise product of the values of each tensor component:
	\begin{align*}
		[A\otimes B]_{i_1, \cdots ,i_r,j_1,\cdots,j_s} := A_{i_1, \cdots ,i_r} \cdot B_{j_1,\cdots,j_s}
	\end{align*}
	In tensor network notation, the tensor product is represented by placing two tensors next to each other. 
\end{definition}
\begin{definition}
	The (partial) \emph{trace} is a joint summation over two indices of a given tensor that have the same dimension. The following example shows the trace operation for a tensor $T$ where the dimensions $d_x$ and $d_y$ are equal:
	\begin{align*}
		&[\tr_{x,y} (A)]_{i_1,\cdots ,i_{x-1}, i_{x+1},\cdots i_r,j_1,\cdots ,j_{y-1}, j_{y+1},\cdots j_s} \\ &= \sum_{\alpha}^{d_x} A_{i_1,\cdots ,i_{x-1},\alpha, i_{x+1},\cdots i_r,j_1,\cdots ,j_{y-1},\alpha, j_{y+1},\cdots j_s}
	\end{align*}
	In tensor network notation, this summation is represented by joining the corresponding tensor legs together:
	\begin{center}
		\begin{tikzpicture}[scale=0.8]
			\node[above] at (0.2-2.5,1.2) {\scriptsize $i$};
			\node[below] at (0.2-2.5,-0.2) {\scriptsize $i$};
			\node[above] at (0.8-2.5,1.2) {\scriptsize $j$};
			\node[below] at (0.8-2.5,-0.2) {\scriptsize $k$};
			\draw (0.2-2.5,1.2) -- (0.2-2.5,-0.2);
			\draw (0.8-2.5,1.2) -- (0.8-2.5,-0.2);
			\draw[fill=white,line width=0.5] (0-2.5, 0) rectangle (1-2.5,1);
			\node at (0.5-2.5,0.5) {$A$};
			\node[left] at (-2.9,0.5) {$\tr_{i}$};
			\node at (-0.2-2.5,0.5) {$\left(\vphantom{\dfrac{\dfrac12}{1}}\right.$};
			\node at (1.2-2.5,0.5) {$\left.\vphantom{\dfrac{\dfrac12}{1}}\right)$};
			\node[left] at (-0.5,0.5) {$=$};
			\node[right] at (-0.6,0.5) {$\sum\limits_{i}^{}$};
			\node[above] at (0.2+0.4,1.2) {\scriptsize $i$};
			\node[below] at (0.2+0.4,-0.2) {\scriptsize $i$};
			\node[above] at (0.8+0.4,1.2) {\scriptsize $j$};
			\node[below] at (0.8+0.4,-0.2) {\scriptsize $k$};
			\draw (0.2+0.4,1.2) -- (0.2+0.4,-0.2);
			\draw (0.8+0.4,1.2) -- (0.8+0.4,-0.2);
			\draw[fill=white,line width=0.5] (0+0.4, 0) rectangle (1+0.4,1);
			\node at (0.5+0.4,0.5) {$A$};
			\node at (-0.2+0.4,0.5) {$\left(\vphantom{\dfrac{\dfrac12}{1}}\right.$};
			\node at (1.2+0.4,0.5) {$\left.\vphantom{\dfrac{\dfrac12}{1}}\right)$};
			\node at (-0.5+2.5,0.5) {$=$};
			\draw (1.+2.5,0.5) ellipse (0.8 and 1.2);
			\draw (0.8+2.5,1.2) -- (0.8+2.5,-0.2);
			\node[above] at (0.8+2.5,1.15) {\scriptsize $j$};
			\node[below] at (0.8+2.5,-0.15) {\scriptsize $k$};
			\draw[fill=white,line width=0.5] (0+2.5, 0) rectangle (1+2.5,1);
			\node at (0.5+2.5,0.5) {$A$};
		\end{tikzpicture}\vspace{4mm}
		\begin{tikzpicture}[scale=0.8]
			\node[above] at (0.2-2.5,1.2) {\scriptsize $i$};
			\node[below] at (0.2-2.5,-0.2) {\scriptsize $i$};
			\node[above] at (0.8-2.5,1.2) {\scriptsize $j$};
			\node[below] at (0.8-2.5,-0.2) {\scriptsize $j$};
			\draw (0.2-2.5,1.2) -- (0.2-2.5,-0.2);
			\draw (0.8-2.5,1.2) -- (0.8-2.5,-0.2);
			\draw[fill=white,line width=0.5] (0-2.5, 0) rectangle (1-2.5,1);
			\node at (0.5-2.5,0.5) {$A$};
			\node[left] at (-2.9,0.5) {$\tr_{}$};
			\node at (-0.2-2.5,0.5) {$\left(\vphantom{\dfrac{\dfrac12}{1}}\right.$};
			\node at (1.2-2.5,0.5) {$\left.\vphantom{\dfrac{\dfrac12}{1}}\right)$};
			\node[left] at (-0.5,0.5) {$=$};
			\node[right] at (-0.6,0.5) {$\sum\limits_{i,j}^{}$};
			\node[above] at (0.2+0.4,1.2) {\scriptsize $i$};
			\node[below] at (0.2+0.4,-0.2) {\scriptsize $i$};
			\node[above] at (0.8+0.4,1.2) {\scriptsize $j$};
			\node[below] at (0.8+0.4,-0.2) {\scriptsize $j$};
			\draw (0.2+0.4,1.2) -- (0.2+0.4,-0.2);
			\draw (0.8+0.4,1.2) -- (0.8+0.4,-0.2);
			\draw[fill=white,line width=0.5] (0+0.4, 0) rectangle (1+0.4,1);
			\node at (0.5+0.4,0.5) {$A$};
			\node at (-0.2+0.4,0.5) {$\left(\vphantom{\dfrac{\dfrac12}{1}}\right.$};
			\node at (1.2+0.4,0.5) {$\left.\vphantom{\dfrac{\dfrac12}{1}}\right)$};
			\node at (-0.5+2.5,0.5) {$=$};
			\draw (1.+2.5,0.5) ellipse (0.8 and 1.2);
			\draw (1.+2.5,0.5) ellipse (0.4 and 0.8);
			\draw[fill=white,line width=0.5] (0+2.5, 0) rectangle (1+2.5,1);
			\node at (0.5+2.5,0.5) {$A$};
		\end{tikzpicture}
	\end{center}	
\end{definition}
\begin{definition}A \emph{contraction} is a tensor product (two tensors are placed next to each other) followed by a trace between corresponding indices of the two tensors. In tensor network notation this can be represented as follows:
	\begin{center}
		\begin{tikzpicture}[scale=0.42]
			\node[left] at (-2,1.2) {$\sum\limits_{i,j}$};
			\draw (0,0) -- (-1.3,1);
			\draw (0,0) -- (1.3,1);
			\draw (0,0) -- (0,-1.);
			\draw[fill=white] (0,0) circle (0.5);
			\draw (0,2.5) -- (-1.3,-1+2.5);
			\draw (0,2.5) -- (1.3,-1+2.5);
			\draw (0,2.5) -- (0,1.+2.5);
			\draw[fill=white] (0,2.5) circle (0.5);
			\node[left] at (-1.3,1.) {\scriptsize $i$};
			\node[right] at (1.3,1.) {\scriptsize $j$};
			\node[left] at (-1.3,1.5) {\scriptsize $i$};
			\node[right] at (1.3,1.5) {\scriptsize $j$};
			\node[right] at (1.7,1.25) {$=$};
			\draw (3.5,1.25) ellipse (0.7 and 1.3);
			\draw (3.5,0) -- (3.5,-1.);
			\draw[fill=white] (3.5,0) circle (0.5);
			\draw (3.5,2.5) -- (3.5,1.+2.5);
			\draw[fill=white] (3.5,2.5) circle (0.5);
		\end{tikzpicture}
	\end{center}
\end{definition}

\section{Tensor networks with $\ell\neq 1$ are partial isometries}
\label{app:PIlneq1}

It can easily be shown, that the tensor networks represelting de Sitter spacetime with $\ell \neq 1$ are also partial isometries. The following calculation shows this for the tensor network for $\ell=2$:\\

\includegraphics[page=37]{figures.pdf}\\

An analogous calculation can be carried out for larger values of $\ell$. 

\section{Matrix representation for the action of isometries on the future boundary of de Sitter spacetime}
\label{app:symmMatrix}
In this appendix we explicitly describe an isogeny of the group $\SL(2,\mathbb{R})$ to isometry group $\SO(1,2)$ of de Sitter spacetime. We further calculate the induced action of an element of $\SL(2,\mathbb{R})$ on null geodesics, and thereby, on the temporal boundaries. 

We commence by first finding an explicit expression for the parameters $s$, $u$ and $v$ defining a null geodesic in terms of the embedding coordinates of the null geodesic (we consider both \emph{anticlockwise} pointing null geodesics $x_\text{ac}(s)$ and \emph{clockwise} pointing null geodesics $x_\text{c}(s)$):
\begin{equation}
	x_\text{ac}(s)
	=\begin{pmatrix}s \\ u + vs \\ v - us\end{pmatrix} \quad \Rightarrow \quad 
	\begin{array}{r l}
		x_{\text{ac},0} &= s\\
		x_{\text{ac},1} &= u + v s\\
		x_{\text{ac},2} &= v - us
	\end{array}\label{eq:acDGL}
\end{equation}
Hence we find
\begin{equation}
	u = \frac{x_{\text{ac},1} - x_{\text{ac},0} x_{\text{ac},2}}{1+x_{\text{ac},0}^2}, \quad \text{and}\quad 
	v = \frac{x_{\text{ac},2} + x_{\text{ac},0} x_{\text{ac},1}}{1+x_{\text{ac},0}^2}. \label{eq:geodesUVanticl}
\end{equation}
Similarly, for clockwise pointing null geodesics
\begin{equation}
	x_\text{c}(s)
	=\begin{pmatrix}s \\ u - vs \\ v + us\end{pmatrix}\quad \Rightarrow \quad 
	\begin{array}{r l}
		x_{\text{c},0} &= s\\
		x_{\text{c},1} &= u - v s\\
		x_{\text{c},2} &= v + us
	\end{array}\label{eq:cDGL}
\end{equation}
we have
\begin{equation}
	u = \frac{x_{\text{c},1} + x_{\text{c},0} x_{\text{c},2}}{1+x_{\text{c},0}^2}, \quad \text{and}\quad 
	v = \frac{x_{\text{c},2} - x_{\text{c},0} x_{\text{c},1}}{1+x_{\text{c},0}^2}.
\end{equation}

We now analyse the action of a spacetime isometry on the future boundary of de Sitter spacetime by exploiting a sporadic $2$-to-$1$ homomorphism \cite{garrett_sporadic_2015}
\begin{align*}
	h : \mathrm{SL}(2,\mathbb{R}) \rightarrow \mathrm{SO}(1,2).
\end{align*}
This homomorphism is constructed via an auxiliary vector space $V$ defined by the space of real-valued $2\times2$-matrices with trace $0$ and symmetric bilinear form given by
\begin{align}\langle x,y\rangle = \tr(x y). \label{eq:BilForm}\end{align} 
We choose a basis of $V$ in terms of Pauli-type matrices:
\begin{align}
	e_1 = \begin{pmatrix} 0 & 1 \\ -1 & 0 \end{pmatrix}, \quad
	e_2 = \begin{pmatrix} 0 & 1 \\ 1 & 0 \end{pmatrix}, \quad
	e_3 = \begin{pmatrix} 1 & 0 \\ 0 & -1 \end{pmatrix}. \label{eq:basisV}
\end{align} 
With respect to this basis the bilinear form has matrix elements $\langle e_j,e_k\rangle$, and is manifestly equivalent to the Minkowski metric with matrix representation
\begin{align}
	\begin{pmatrix} -1 & 0 & 0 \\ 0 & 1 & 0 \\ 0 & 0 & 1 \end{pmatrix},
\end{align} 
allowing us to identify Minkowski spacetime $\mathbb{R}^{1,2}$ with $V$. 
The action of $\SL(2,\mathbb{R})$ on the space $V$ is defined by 
\begin{align}
	M \mapsto g\cdot M = g M g^{-1},
\end{align}
where $g\in \SL(2,\mathbb{R})$. 
This action preserves the bilinear form defined in Eq.~(\ref{eq:BilForm}), because the trace is cyclic and $g$ multiplied with its inverse yields the identity:
\begin{align}\begin{split}
		\langle M,N\rangle &= \frac{1}{2}\tr(M N) \mapsto \,\frac{1}{2}\tr(g M g^{-1} g N g^{-1}) \\&= \langle g\cdot M,g\cdot N\rangle
		= \frac{1}{2}\tr(M N) =	\langle M,N\rangle.
\end{split}\end{align}
Thus, according to this homomorphism, $\SL(2,\mathbb R)$ is identified with a copy of the special orthogonal group $\SO(1, 2)$.
The kernel of this action is determined by the set of all matrices $g$ for which 
\begin{align}
	g\cdot M = M \quad \Leftrightarrow \quad g M g^{-1} = M.
\end{align}
Thus the kernel is given by the set of all $g\in \SL(2,\mathbb{R})$ commuting with all elements $M\in V$. This is only true for the elements $\{\pm \mathbb{I}\}$, so we have a double covering.

We now explicitly calculate the matrix elements of the transformation $h\in \SO(1, 2)$ corresponding to a given element $g\in \SL(2,\mathbb{R})$ with matrix representation
\begin{align*}g=\begin{pmatrix} a & b \\ c & d \end{pmatrix}.\end{align*} 
Any $M\in V$ can be expressed as a linear combination of the basis elements $e_j$.
In order to express the function $h$ with regard to this basis, we exploit the homomorphism:
\begin{equation}
	e_j \mapsto g \sigma_i g^{-1} \equiv h_{j 1}\, e_1 + h_{j 2}\, e_2 + h_{j 3}\, e_3 
\end{equation}
This leads to the following matrix representation: 
\begin{widetext}
	\begin{equation}
		h(g)=
		\begin{pmatrix}
			\frac{1}{2} \left(a^2+b^2+c^2+d^2\right) \hspace*{5mm}& \frac{1}{2} \left(a^2-b^2+c^2-d^2\right) \hspace*{5mm}& -a b-c d \\[1.5ex]
			\frac{1}{2} \left(a^2+b^2-c^2-d^2\right) & \frac{1}{2} \left(a^2-b^2-c^2+d^2\right) & c d-a b \\[1.5ex]
			-a c-b d & b d-a c & b c+a d \label{eq:actionMatrix}
		\end{pmatrix}.
	\end{equation}
	In order to analyse the action of the isometry group $\SO(1,2)$, we use that an isometry is completely specified by its action on the null geodesics. 
	The action on a null geodesic can be obtained by applying the matrix representation for $h$ to a null geodesic $x(s)$:
	\begin{align}
		x(s)= \begin{pmatrix} s \\ u+v s \\ v-u s \end{pmatrix} \quad \rightarrow \quad x'(s')= h(g) x(s) = \begin{pmatrix}
			s' \\ u' + v' s' \\ v' - u' s'
		\end{pmatrix} = \begin{pmatrix}
			x_0' \\ x_1' \\x_2'
		\end{pmatrix}
	\end{align}
	Thus
	\begin{align*}
		x' =  \begin{pmatrix}
			\frac{1}{2} \left((s v+u) \left(a^2-b^2+c^2-d^2\right)+s \left(a^2+b^2+c^2+d^2\right)\right) + (s u-v) (a b+c d)\\[1.5ex]
			\frac{1}{2} \left((s v+u) \left(a^2-b^2-c^2+d^2\right)+s \left(a^2+b^2-c^2-d^2\right)\right)+(v-s u) (c d-a b) \\[1.5ex]
			(v-s u) (a d+b c)+(s v+u) (b d-a c)-s (a c+b d) \\
		\end{pmatrix}.
	\end{align*}
	We are interested in the symmetry action the temporal boundaries, from which we infer the transformation rules of the tessellation and therefore our holographic network. By considering the limit $s'\rightarrow \infty$ we obtain the action of $h(g)$ on the temporal future boundary $\mathcal{I}^+$: 
	\begin{align}
		\dfrac{x'(s)}{s'} = \begin{pmatrix}
			1 \\[0.5ex] \dfrac{u'}{s'} + v' \\[1.5ex] \dfrac{v'}{s'} - u'
		\end{pmatrix}  \underset{s'\rightarrow\infty}{\longrightarrow} \left(\begin{array}{c}
			1\\v'\\-u'
		\end{array}\right).\end{align}
	With this expression for the symmetry action on the future boundary of de Sitter spacetime, the parameters $u'$ and $v'$ can easily be calculated using the relation derived in Eq.~(\ref{eq:geodesUVanticl}). The new parameters $u'$ and $v'$ satisfy the condition $u'^2+v'^2 =1$.
	{
		\begin{subequations}\begin{align}
				u' =\,&  \frac{x_1' - x_0' x_2'}{1+(x_0')^2}
				=\,\frac{-(-a c u+a d v+b c v+b d u) \left(u \left(a^2-b^2+c^2-d^2\right)-2 v (a b+c d)\right)}{2 \left(\frac{1}{4} \left(u \left(a^2-b^2+c^2-d^2\right)-2 v (a b+c d)\right)^2+1\right)}\nonumber \\&+ 
				\frac{u \left(a^2-b^2-c^2+d^2\right)+v (2 c d-2 a b)}{2 \left(\frac{1}{4} \left(u \left(a^2-b^2+c^2-d^2\right)-2 v (a b+c d)\right)^2+1\right)}\\
				v' =\,& \frac{x_2' + x_0' x_1'}{1+(x_0')^2}
				=\,\frac{\frac{1}{4} \left(u \left(a^2-b^2-c^2+d^2\right)-2 a b v+2 c d v\right) \left(u \left(a^2-b^2+c^2-d^2\right)-2 v (a b+c d)\right)}{\frac{1}{4} \left(u \left(a^2-b^2+c^2-d^2\right)-2 v (a b+c d)\right)^2+1} \nonumber \\&
				+ \frac{u (b d-a c)+v (2 a d-1)}{\frac{1}{4} \left(u \left(a^2-b^2+c^2-d^2\right)-2 v (a b+c d)\right)^2+1}.
			\end{align}\label{eq:uvMatrix}\end{subequations}
	}
\end{widetext}

\end{appendix}

\bibliographystyle{apsrev4-2}
\bibliography{literature}

\end{document}